\documentclass[twocolumn,trackchanges]{aastex7}

\usepackage{amsmath}
\usepackage{ragged2e}
\usepackage{booktabs}
\usepackage{subfig}
\usepackage{caption}
\usepackage[english]{babel}
\PassOptionsToPackage{disable}{longtable} 
\usepackage{adjustbox} 
\usepackage{graphicx}  
\usepackage{tabularx}

\begin{document}

\title{UHECRs Propagation and their Multimessengers: Upper limits and the Impact of the Extragalactic Magnetic Field}

\author[0009-0003-4952-7292]{Rodrigo Sasse}
\affiliation{Programa de Pós-graduação em Física \& Departamento de Física,\\ Universidade Estadual de Londrina (UEL), Rodovia Celso Garcia Cid Km 380, 86057-970 Londrina, PR, Brazil}
\affiliation{Programa de Pós-Graduação em Física Aplicada,\\ Universidade Federal da Integração Latino-Americana, 85867-670, Foz do Igua\c{c}u, PR, Brazil}
\email[show]{rodrigo.sasse1@uel.br}

\author[0000-0002-2893-7589]{Rubens Jr. Costa}
\affiliation{Programa de Pós-graduação em Física e Astronomia, \\ Universidade Tecnológica do Paraná (UTFPR), 80230-901, Curitiba, PR, Brazil}
\email[show]{rubensp@utfpr.edu.br}

\author[0000-0002-0978-6316]{Adriel G.B. Mocellin}
\affiliation{Programa de Pós-graduação em Física e Astronomia, \\ Universidade Tecnológica do Paraná (UTFPR), 80230-901, Curitiba, PR, Brazil}
\affiliation{Department of Physics, Colorado School of Mines (CSM), 1500 Illinois St, Golden-CO, 80401, USA}
\email[show]{\\ agbartzmocellin@mines.edu} 

\author[0000-0003-3588-2587]{Carlos H. Coimbra-Araujo}
\affiliation{Programa de Pós-Graduação em Física Aplicada,\\ Universidade Federal da Integração Latino-Americana, 85867-670, Foz do Igua\c{c}u, PR, Brazil}
\affiliation{Departamento de Engenharias e Exatas,\\ Universidade Federal do Paraná (UFPR), Pioneiro, 2153, 85950-000 Palotina, PR, Brazil}
\email[show]{carlos.coimbra@ufpr.br}

\author[0000-0002-6463-2272]{R. C. Dos Anjos}
\affiliation{Programa de Pós-graduação em Física \& Departamento de Física,\\ Universidade Estadual de Londrina (UEL), Rodovia Celso Garcia Cid Km 380, 86057-970 Londrina, PR, Brazil}
\affiliation{Max-Planck-Institut für Kernphysik, Saupfercheckweg 1, D-69117 Heidelberg, Germany}
\affiliation{Departamento de Engenharias e Exatas,\\ Universidade Federal do Paraná (UFPR), Pioneiro, 2153, 85950-000 Palotina, PR, Brazil}
\affiliation{N\'ucleo de Astrofísica e Cosmologia (Cosmo-Ufes), Universidade Federal do Esp\'irito Santo, 29075--910, Vit\'oria, ES, Brazil}
\affiliation{Programa de Pós-graduação em Física e Astronomia, \\ Universidade Tecnológica do Paraná (UTFPR), 80230-901, Curitiba, PR, Brazil}
\affiliation{Programa de Pós-Graduação em Física Aplicada,\\ Universidade Federal da Integração Latino-Americana, 85867-670, Foz do Igua\c{c}u, PR, Brazil}
\email[show]{\\ritacassia@ufpr.br} 



\begin{abstract}
The detection of high-energy astrophysical multimessengers establishes a connection between ultra-high-energy cosmic rays (UHECRs) and powerful cosmic accelerators. Interactions of UHECRs with radiation fields and interstellar matter generate very-high-energy (VHE) gamma rays and neutrinos, making them key components in the multimessenger framework. This study examines the cosmogenic gamma-ray and neutrino fluxes resulting from UHECR propagation in starburst galaxies with supernova remnants, with a particular focus on NGC 1068, a well-established high-energy neutrino source. Using extragalactic simulations, we calculate the upper limit on cosmic-ray luminosity, applying upper limits on gamma-ray fluxes derived from observations by H.E.S.S. and MAGIC observatories.
Our analysis incorporates the effects of both extragalactic and galactic magnetic fields on particle propagation, constraining the maximum extragalactic magnetic field (EGMF) intensity to $10^{-14}~\mathrm{G}$ to ensure that at least 90\% of injected UHECRs successfully reach Earth. The results provide upper limits on gamma-ray and neutrino fluxes, estimates of UHECR luminosity for individual sources, and predictions for the detection capabilities of the Cherenkov Telescope Array Observatory regarding gamma-ray emission from NGC 1068. Combining gamma-ray, neutrino, and UHECR observations reinforces the importance of multimessenger approaches in understanding the nature of high-energy astrophysical sources and their role in cosmic-ray acceleration.
\end{abstract}

\keywords{\uat{Gamma-rays}{637}; \uat{Gamma-ray observatories}{632}; \uat{Cosmic rays}{329}; \uat{Neutrino Astro-nomy}{1100}}


\section{Introduction} \label{sec:intro}

Ultra-high-energy cosmic rays (UHECRs), predominantly atomic nuclei with $\sim 10^{18} {\rm eV}$, rank among the most energetic particles in the universe. However, identifying their sources remains a challenge in astrophysics. The low number of these particles and deflections in their trajectories caused by magnetic fields make it difficult to determine their origins. This limits our ability to study individual UHECR sources and hinders our understanding of the astrophysical accelerators responsible for their production \citep{arrival2017, 2018ApJ...853L..29A}.

A multimessenger approach is the most effective way to address these challenges. In particular, interactions of UHECRs with radiation fields and interstellar matter through non-thermal processes can generate very high-energy (VHE, $10^{11}\ \mathrm{eV} \lesssim E \lesssim 10^{14}\ \mathrm{eV}$) gamma rays and neutrinos. These secondary cosmic messengers are fundamental components of the multimessenger framework. These interactions can occur within the source environment, such as in active galactic nuclei (AGNs) or starburst galaxies, where dense matter or intense radiation fields lead to hadronic processes, and those taking place during extragalactic propagation, where UHECRs interact with background photon fields like the cosmic microwave background (CMB) and extragalactic background light (EBL). Hadronic interactions, such as proton-proton ($pp$) and and proton-nucleus ($pA$) collisions, produce mesons following the astrophysical beam dump mechanism: p + p $\rightarrow \pi^{\pm}, \pi^{0}$, K$^{\pm}$, K$^{0}$, p, n. These mesons subsequently decay into secondary particles gamma rays, neutrinos, and leptons. A second mechanism that generates secondary mesons involves high-energy protons interacting with low-energy photons from the cosmic microwave background (CMB)~\citep{ANCHORDOQUI20191}. This interaction takes place via photoproduction, predominantly through the resonant production $\Delta^{+}$ of the hadron in the center-of-mass frame, which subsequently undergoes rapid decay into the following products:
\begin{equation}\label{p1}
    \begin{aligned}
        \mathrm{p} + \gamma_{\mathrm{CMB}} \rightarrow \Delta^{+} 
&\rightarrow \pi^{0} \mathrm{p} \rightarrow \gamma \gamma, \\
        &\phantom{\rightarrow \pi^{0} \mathrm{p}\ } \rightarrow e^{-} e^{+} \gamma,\\
        &\rightarrow \pi^{+} \mathrm{n}, 
    \end{aligned}
\end{equation}
To produce pions via photoproduction, a proton must exceed the energy threshold of $E_{\mathrm th} \approx 280\ \mathrm{MeV}$. The decay of neutral pions results in secondary gamma rays, representing one of the most significant outcomes of this interaction. In this process, pions serve as intermediaries, facilitating the protons' kinetic energy conversion into gamma rays. Meanwhile, charged pions undergo decay into muons, which subsequently decay further, following the decay chain:
\begin{equation}\label{p2}
\begin{aligned}
    \pi^{\pm} \rightarrow &\mu^{\pm} + \nu_{\mu}(\overline{\nu}_{\mu}), \\
    &\mu^{\pm} \rightarrow e^{\pm} + \nu_{e}(\overline{\nu}_{e}) + \overline{\nu}_{\mu}(\nu_{\mu}).
\end{aligned}
\end{equation}

Several factors influence the neutrino flux, including the cosmological evolution of cosmic-ray sources, their luminosity, and the fraction of proton energy converted into charged pions. These pions subsequently decay into neutrinos, which is crucial in shaping the expected flux. In resonant photoproduction, approximately 28\% of the nucleon's energy is transferred to charged pions at the source  ($\epsilon_\pi \approx 0.28$)~\citep{ANCHORDOQUI20191}. Recent observations, such as the detection of a UHE cosmic neutrino with KM3NeT, provide valuable insights into these interactions, reinforcing the importance of multimessenger studies in understanding the origin and propagation of UHECRs~\citep{km3net_2025}.

When photopion interactions dominate the intergalactic medium via the resonance $\Delta^{+}$, neutral pions decay into gamma rays, triggering an electromagnetic cascade in radiation fields, as described in Equation~\ref{p1}. This cascade is followed by inverse Compton scattering, producing gamma rays in the GeV–TeV energy range. Meanwhile, the decay of charged pions generates neutrinos, electrons, and positrons that contribute significantly to inverse Compton scattering or synchrotron radiation, further shaping the observed electromagnetic signatures. The intensity of cosmogenic neutrinos is strongly influenced by the energy loss of UHECRs as they propagate toward Earth. In particular, neutrinos carry approximately 75\% of the $\pi^{+}$ energy, underscoring the strong correlation between the energy distributions of cosmogenic neutrinos and gamma rays. For a proton-dominated UHECR composition, the peak in neutrino flux is expected to occur between $10^{9}$ and $10^{10}\mathrm{\ GeV}$, whereas heavier nuclei shift this peak to lower energies, around $10^{8.7}\mathrm{\ GeV}$. Understanding these connections is essential for advancing our knowledge of high-energy cosmic phenomena and refining multimessenger astrophysics models \citep[for recent reviews, see][]{2018pma..book.....S, ANCHORDOQUI20191, 2022hxga.book..104S}.

Primarily through pion decay, secondary fluxes of multimessenger particles are generated, forming the basis for the mathematical framework discussed in Section \ref{section2}. The flux of VHE gamma rays and cosmogenic neutrinos serves as a crucial probe of the source luminosity of the UHECR, particularly at energies exceeding $10^{18}\ \mathrm{eV}$. By analyzing these gamma-ray and neutrino emissions, valuable information about UHECR sources can be indirectly obtained, offering insights into the mechanisms governing particle acceleration and propagation in extreme astrophysical environments (\citealt{2025Univ...11...22L, 2022A&A...658L...6D, Das:2020hev, universe6020030,  Aloisio_2017, ALOISIO201373}).

In this paper, we utilize cosmogenic gamma rays and neutrinos to estimate the UHECR luminosity of point-like sources. We refine the method proposed by (\citealt{Supanitsky_2013, Anjos_2014}), incorporating updated observational data from the Pierre Auger Observatory (PAO)~\citep{PhysRevLett.125.121106}. In addition, we improve the approach by accounting for the impact of extragalactic magnetic fields (EGMFs). For NGC 1068, a prominent source of high-energy neutrinos \citep{2022Sci...378..538I}, we analyze its observability with the Cherenkov Telescope Array Observatory (CTAO) \citep{2019scta.book.....C}, assessing its potential for gamma-ray detection. The structure of this paper is as follows: Section~\ref{section2} presents the details of the mathematical approach and the simulation framework, along with analyses of multimessenger fluxes under different scenarios. Section~\ref{section3} discusses the source selection criteria for applying the method and provides upper limits (UL) on UHECR luminosity for individual sources. Section~\ref{sec:ctao} evaluates the performance of the CTAO in observing NGC 1068. Finally, Section~\ref{conclusions} summarizes this study's key findings and conclusions.

\section{Modeling GeV-TeV gamma-ray emissions and UHECR Luminosity} \label{section2}

This Section presents the method based on the approaches developed in \cite{Supanitsky_2013, Anjos_2014}, which relate the integral gamma-ray flux to the UHECR luminosity of astrophysical sources. This study specifically focuses on galaxies with GeV-TeV observations that constrain their gamma-ray flux integrals, allowing for a multimessenger analysis of their cosmic-ray emissions. By measuring the UL on the integral GeV-TeV gamma-ray flux for a given source, the corresponding UL on its UHECR luminosity can be determined. For clarity, we present a refined description of the methodology to improve comprehensibility.

When direct gamma-ray flux observations are available, the detected flux may originate from both gamma rays emitted directly by the source and those produced during the propagation of UHECRs. This introduces complexity in evaluating the method's conservativeness, as the inferred UHECR luminosity may not fully represent the total source's luminosity. Nevertheless, this approach remains applicable to any UL on the integral GeV-TeV gamma-ray flux derived from ground-based or space-based instruments, providing a framework for constraining UHECR source properties.

Following the formalism in \citep{Supanitsky_2013, Anjos_2014}, we assume a source with a cosmic-ray luminosity $L_{\rm CR}$\footnote{In this work, the subscripts ${\rm CR}$, $\gamma$, and $\nu$ are used to denote quantities related to cosmic rays, gamma rays, and neutrinos, respectively.} injecting particles with an exponential cutoff power-law spectrum: $ \propto  E^{-\alpha} e^{- E/ E_{\rm{cut}}}$, where $\alpha$ is the spectral index and $ E_{\rm cut}$ is the cutoff energy. The corresponding injection spectrum, defined per unit solid angle $\Omega$ and time $t$, is expressed as: 
\begin{equation}\label{eq:spec}
    \frac{dN}{dE \, dt \, d\Omega} = \frac{L_{\mathrm{CR}}}{\langle E \rangle_{0}} P_{\mathrm{CR}}^{0}(E),
\end{equation} 
where $N$ is the number of injected particles, $P_{\mathrm CR}^{0}$ is the normalized energy distribution, and $\langle \rm E \rangle_{0}$ is the average particle energy, defined as:
\begin{equation}
    \langle E \rangle_{0} = \int_{E_{\mathrm{min}}}^{\infty} dE  E P_{\mathrm{CR}}^{0}(E) = \frac{\int_{E_{\mathrm{min}}}^{\infty} dE \, E^{-\alpha+1} e^{-E/E_{\mathrm{{cut}}}}}{\int_{E_{\mathrm{min}}}^{\infty} dE E^{-\alpha} e^{-E/E_\mathrm{cut}}},
\end{equation}
where $ E_{\rm min}$ is the minimum energy of injected particles. For an isotropically emitting source located at a comoving distance $D$ from Earth and considering energy losses during propagation, the observed cosmic-ray flux $I_{\mathrm CR}$ is given by:
\begin{equation}\label{eq:fluxCR}
  I_{\mathrm{CR}}(E) = \frac{L_{\mathrm{CR}} W(\hat{n})}{4 \pi D^{2} (1 + z) \langle E \rangle_{0}} K_{\mathrm{CR}} P_{\mathrm{CR}}(E).
\end{equation}
Here, $z$ denotes the source redshift, $P_{\mathrm{CR}}$ represents the energy distribution of the arriving particles and $K_{\mathrm{CR}}$ is the normalization factor, defining the number of cosmic rays relative to the injected particles. The term $W(\hat{n})$ represents the source weight, incorporating both its position in the sky, $\hat{n}$, and the observational exposure of the facility. 

Due to energy-loss processes during cosmic-ray propagation, the cosmic-ray source also produces a secondary gamma-ray flux that intrinsically correlates to the cosmic-ray luminosity. Consequently, the observed gamma-ray flux, $I_{\gamma}$, can be expressed in terms of the corresponding spectral distribution $P_{\gamma}(E_{\gamma}$),  and normalization factor $K_{\gamma}$, the number of gamma rays relative to injected particles:
\begin{equation}\label{eq:gamma}
     I_{\gamma}(E_{\gamma}) = \frac{L_{\mathrm{CR}}}{4 \pi D^{2} (1 + z) \langle E \rangle_{0}} K_{\gamma} P_{\gamma}(E_{\gamma}),
\end{equation}

The primary mechanisms contributing to gamma-ray production during cosmic-ray propagation include pion photoproduction, pair production, and electron-positron annihilation (\citealt{HEITER201839, ANCHORDOQUI20191}).

The UL on cosmic-ray luminosity $L_{\mathrm CR}^{\mathrm UL}$ for an individual source can be derived from the UL on the integral gamma-ray flux, $ I_{\gamma}^{\mathrm{UL}}$, measured by an observatory at a given energy threshold, $E_{\gamma}^{\mathrm{th}}$, and confidence level (CL):
\begin{equation} \label{eq:LCR}
    L_{\mathrm{CR}}^{\mathrm{UL}} = \frac{4 \pi D^{2} (1 + z) \langle E \rangle_{0}}{K_{\gamma}\int_{E_{\gamma}^{\mathrm{th}}}^{\infty} dE \, P_{\gamma}(E_{\gamma})} I_{\gamma}^{\mathrm{UL}}(> E_{\gamma}^{\mathrm{th}}),
\end{equation}
This equation provides a direct relation between the gamma-ray flux constraints and the UHECR luminosity, offering insights into potential sources of high-energy cosmic rays \citep[for detailed reviews, see][]{Supanitsky_2013, Anjos_2014, Sasse_2021}.

To apply this method, it is necessary to consider the flux UL of gamma-ray, the observatory’s field of view, and exposure. This enables the determination of $\ I_{\gamma}^{\mathrm{UHECR}}$, which acts as a key parameter in assessing the applicability of the method for estimating cosmic-ray luminosity constraints. For an individual cosmic-ray source, two critical factors must be considered: the number of detected events in each energy bin of the cosmic-ray spectrum measured by the Observatory and the source's exposure to the observatory, as referenced in \citep{Anjos_2014}. The exposure is determined using Poisson statistics applied to the cosmic-ray spectrum from the PAO \citep{2011APh....34..368A}. The resulting spectrum provides ULs on the flux for each energy bin, as described by:
\begin{equation}\label{exposure}
   \jmath_{i}^{ \mathrm{UL}} = \frac{\mu_{i}^{ \mathrm{UL}}}{\sigma_{i}}.
\end{equation}
In Equation \ref{exposure}, $\mu_{i}$ represents the UL derived from the number of events in each energy bin of the cosmic-ray spectrum and $\sigma_{i}$ accounts for the source exposure. This calculation is performed at 95\% CL. By applying this methodology, the UL on the integral gamma-ray flux in the GeV–TeV range can be determined as a function of the source distance, denoted as $I_{\gamma}^{\mathrm{UHECR}}$.

The methodology used to derive upper limits on the UHECR luminosity follows a two-step procedure that combines propagation simulations with published gamma-ray upper limits from multiple observatories. First, we establish a selection criterion to identify the sources from which  gamma-ray ULs are derived. In this way, we use the CRPropa3 code to simulate the theoretical secondary gamma-ray flux ($I_{\gamma}^{\mathrm{UHECR}}$) that UHECRs would produce under composition and spectral scenarios. A source is then selected for further analysis only if its published gamma-ray UL, is lower than this theoretically predicted flux. This first analysis ensures that the UL derived from our model can genuinely be restrictive enough to the selected sources. Second, for the subset of sources that meet this criterion, we calculate the final results and derive the UL on the UHECR luminosity ($L_{\mathrm{CR}}^{\mathrm{UL}}$). This framework ensures that our results are derived only from sources where the multimessenger connection is meaningfully constrained by current observational capabilities.

\subsection{The Simulation Framework}

The leading software utilized for cosmic-ray propagation simulations in this study is CRPropa3, a widely adopted tool within the UHECR research community. Currently, in its third version, CRPropa3 features multiple modules that enable the simulation of diverse cosmic-ray propagation scenarios. Developed in C++, it integrates Python as an interface via the SWIG tool, allowing seamless execution of Python commands within its modules. As open-source and freely accessible software, it is continuously updated with new tools, improved propagation models, and enhanced functionalities to support the latest cosmic-ray physics~\citep{Alves_Batista_2022}.

Figure \hyperlink{multimessenger}{1} illustrates the 1D simulation framework applied to the comoving distance of NGC 1068 at 14 Mpc from Earth, following a power law with an exponential cutoff, characterized by a spectral index of $\alpha = 2.0$ and a cutoff energy of $E_{\mathrm{cut}} = Z \times 10^{20} \ \mathrm{eV}$. The simulated UHECR spectrum was normalized based on the UL of the energy spectrum observed by the PAO at 95\% CL. The arrow in Figure \hyperlink{multimessenger}{1} marks the UL of the flux measured by the PAO, which was used as a reference to normalize the simulated cosmic-ray spectrum. Following this normalization, the corresponding secondary GeV–TeV gamma-ray and cosmogenic neutrino spectra were adjusted accordingly. This procedure was systematically repeated for each simulated source, accounting for distance variations. In Figure \hyperlink{multimessenger}{1} the blue point represents the flux required for one expected event after track selection within the central 90\% neutrino energy range of KM3-230213A~\citep{km3net_2025}. The purple and pink shaded regions show the contours 68\% CL for IceCube single power-law fits, with the corresponding piecewise fits indicated by purple and pink points ~\citep{Abbasi_2022,PhysRevD.104.022002}. The red cross marks the IceCube Glashow resonance event~\citep{glashow}. The dots denote upper limits of ANTARES (95\% CL)~\citep{Albert_2024}, Pierre Auger (90\% CL)~\citep{AbdulHalim:2023SN}, and IceCube (90\% CL)~\citep{PhysRevD.98.062003}. The light blue shaded regions illustrate the best-fit time-integrated astrophysical neutrino flux, derived from 10 years of IceCube data~\cite{anchordoqui2021highenergyneutrinosngc1068}.
 
\hypertarget{multimessenger}{\begin{figure}[!ht]
\centering
\includegraphics[width=0.5\textwidth]{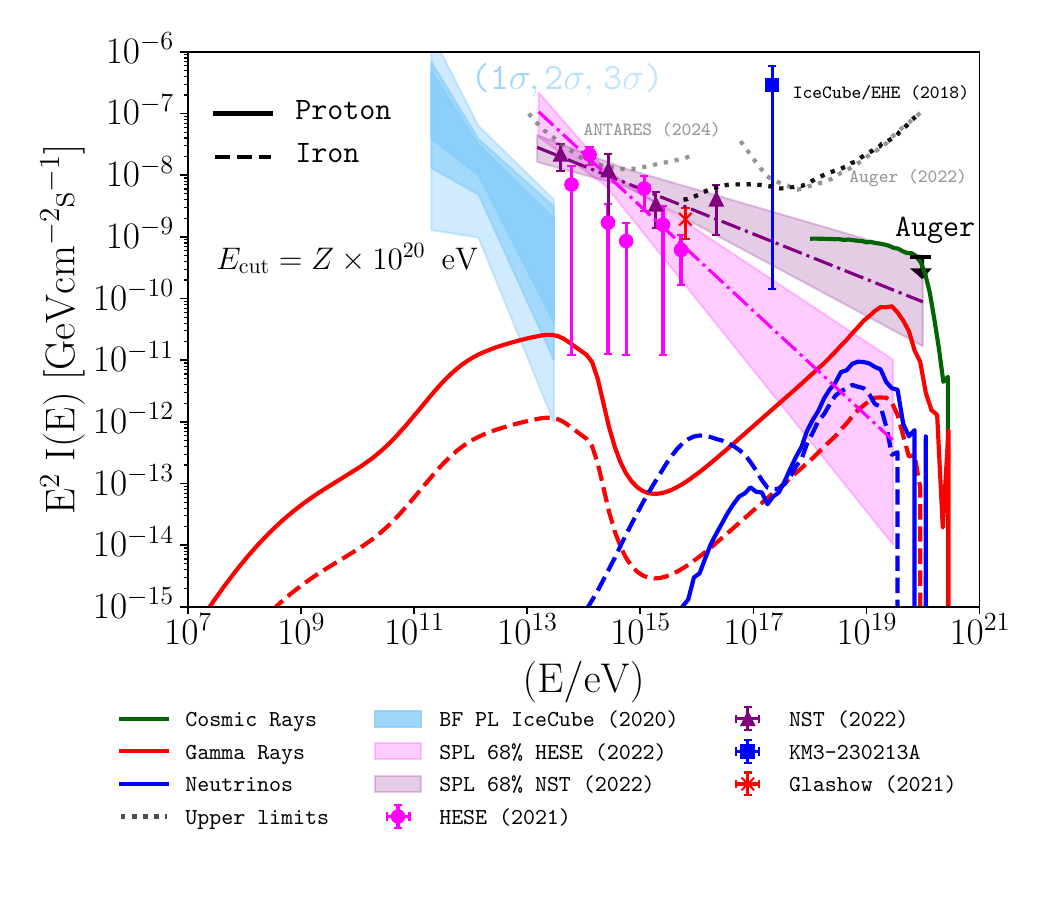}
\caption{Cosmogenic multimessenger fluxes simulated using CRPropa3~\citep{Alves_Batista_2022}. The green curve shows the cosmic-ray spectrum injected at the location of NGC 1068, at a comoving distance of 14 Mpc from Earth. The red curves depict the secondary photon fluxes produced during the propagation of UHECRs, whereas the blue curves represent the resulting cosmogenic neutrino flux. The solid lines correspond to a proton composition for the injected particles, and the dashed lines represent an iron composition scenario. Additionally, the plot includes the UL of the flux measured by the PAO, which serves as a reference for normalizing the simulated cosmic-ray spectrum~\citep{PhysRevLett.125.121106}. This normalization was also applied to the gamma-ray and neutrino fluxes to ensure consistency across multimessenger predictions. See text for details.}
\end{figure}}

In the simulations, the spectrum of injected particles escaping from each extragalactic source is modeled as a composition of contributions from multiple nuclear species. The source spectrum follows a power-law distribution with a broken exponential rigidity cutoff, as described in Equation~\ref{eq:spec}. To reflect a broad range of astrophysical scenarios, we adopt negative spectral indices in the range $\alpha = 2.0$ to $3.0$, consistent with recent UHECR spectrum and composition fits~\cite{2023JCAP...05..024A}. This choice is further motivated by models of powerful, point-like extragalactic sources, such as AGNs and starburst galaxies, where extreme acceleration conditions may allow cosmic rays to reach ultra-high energies up to $E_{\mathrm{max}} = Z \times 100~\mathrm{EeV}$, with $\rm Z$ being the atomic number of the nucleus. We assume a minimum injection energy of $E_{\mathrm{min}} = 10^{18}~\mathrm{eV}$, in line with the energy range relevant to cosmogenic secondary production. The injected UHECRs particles are distributed isotropically. To model cosmic-ray propagation, we performed simulations using a one-dimensional approximation, simulating  $10^7$ particles for each source distance. Additionally, to account for the influence of the EGMF, we employed a four-dimensional simulation incorporating a Kolmogorov spectrum of magnetic turbulence \citep{2022A&A...658L...6D}, ensuring a realistic representation of stochastic deflections during propagation. The different nuclear mass compositions considered in the simulations are summarized in Table~\ref{tab:1}, which presents four mixed compositions based on several nuclear mass models (\citealt{2016PhLB..762..288A, 2016arXiv160403637T, 2023EPJWC.28302013G}).
\begin{table}[h!]
\centering
\renewcommand{\arraystretch}{1.0}
\begin{tabular}{c|ccccc}
\hline
\textbf{Model} & \textbf{H(\%)} & \textbf{He(\%)} & \textbf{N(\%)} & \textbf{Si(\%)} & \textbf{Fe(\%)} \\
\hline
1 & 90  & 10  &  0  &  0  &  0  \\
2 & 90  &  5  &  5  &  0  &  0  \\
3 & 97  &  1  &  1  &  1  &  0  \\
4 & 96  &  1  &  1  &  1  &  1  \\ 
\hline
\end{tabular}
\caption{Mass fractions ($f_{A}(\%)$) of different nuclear species (H, He, N, Si, Fe) for various cosmic-ray composition models.}
\label{tab:1}
\end{table}

The interpretation of the physical mass composition remains uncertain, as it depends on the choice of hadronic interaction models, such as EPOS-LHC, QGSJET-II-04, and Sibyll 2.1 (\citealt{de2018testing, 2021Univ....7..321A}). Our approach (Table~\ref{tab:1}) is consistent with the findings of the Auger Collaboration, whose results indicate a transition toward a heavier cosmic-ray component at energies around $\sim 10^{18.5}\ {\rm eV}$ (\citealt{2024JCAP...01..022A, 2014PhRvD..90l2005A}). This composition model will be used primarily to estimate the UHECR luminosity of individual sources, as detailed in Section~\ref{section3}.

\subsection{Updated Results}\label{updatedresults}

With the release of the updated version of CRPropa3 and a decade since the original studies, it is pertinent to revisit and update some of the results presented by \citealt{Supanitsky_2013, Anjos_2014}. Figure 2 presents the dependence of the integral upper limit (UL) of the gamma-ray flux, $ I^{\rm{UHECR}}_{\gamma}$ ($> E_{\rm th}$), on different parameters relevant to cosmic-ray propagation and multimessenger studies. These ULs are derived using data from the PAO at a 95\% CL, providing information on how different astrophysical and observational factors influence the expected gamma-ray flux from UHECR sources.

\begin{figure}[htbp!]
    \centering
    \begin{minipage}{0.47\textwidth} 
        \centering
        \includegraphics[width=\textwidth]{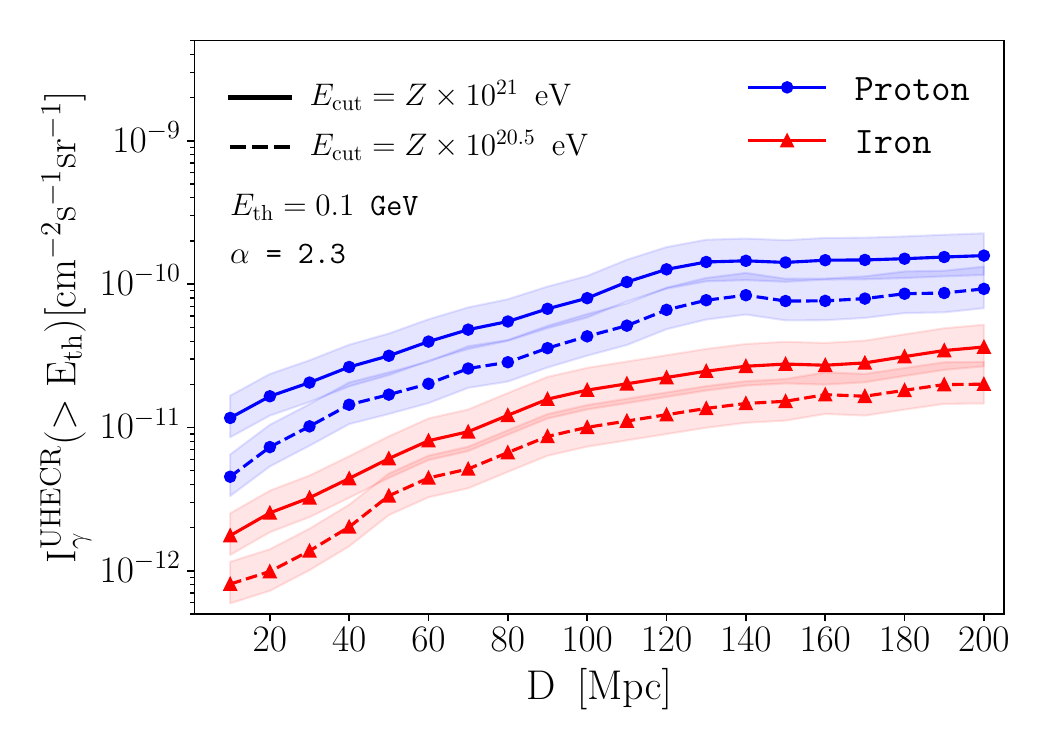} 
        (a) $ E_{\rm cut}$ study
    \end{minipage}
    \begin{minipage}{0.47\textwidth}
        \centering
        \includegraphics[width=\textwidth]{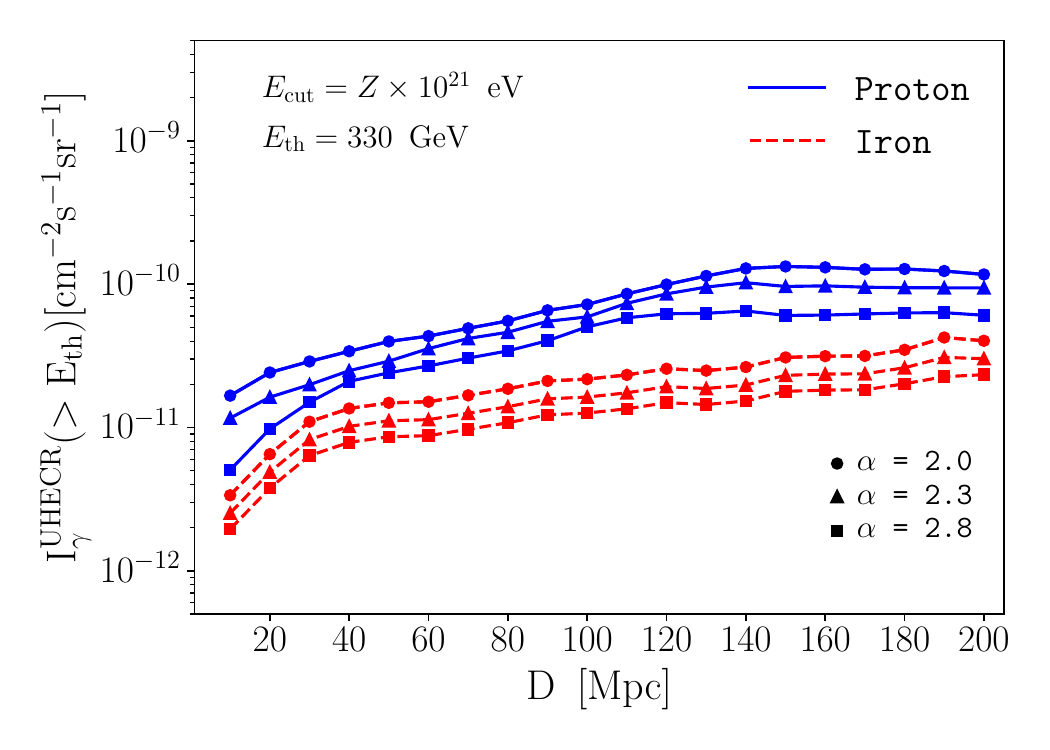}
        (b) $\alpha$ study
    \end{minipage}
    \hfill
    \begin{minipage}{0.47\textwidth}
        \centering
        \includegraphics[width=\textwidth]{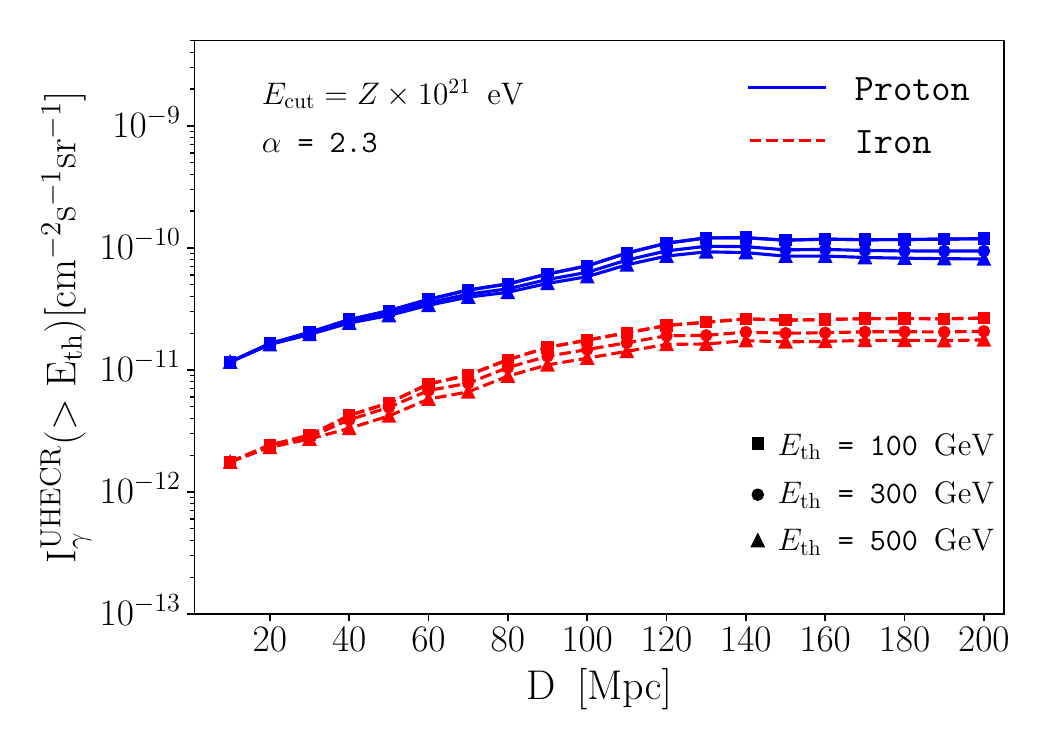}
        (c) $ E_{\rm th}$ study
    \end{minipage}
\caption{Upper limits (ULs) on the integral gamma-ray flux, 
$ I^{\mathrm{UHECR}}_{\gamma}(> E_{\mathrm{th}})$, as a function of the source distance,  
calculated using the UL on the flux observed by the Pierre Auger Observatory  
at the 95\% confidence level (CL) for a given energy threshold $ E_{\mathrm{th}}$.  
(a) Dependence on the cutoff energy $ E_{\mathrm{cut}}$, with fixed parameters  
$\alpha = 2.3$ and $ E_{\mathrm{th}} = 0.1 \ \mathrm{GeV}$.  
(b) Dependence on the spectral index $\alpha$, for fixed parameters  
$ E_{\mathrm{cut}} = Z \times 10^{21} \ \mathrm{eV}$ and  
$ E_{\mathrm{th}} = 330 \ \mathrm{GeV}$.  
(c) Dependence on the energy threshold $ E_{\mathrm{th}}$,  
for fixed parameters $ E_{\mathrm{cut}} = Z \times 10^{21} \ \mathrm{eV}$  
and $\alpha = 2.3$.}
    \label{fig:1dgamma}
\end{figure}

Figure~\ref{fig:1dgamma}-a explores the effect of the cutoff energy $E_{\rm th}$ on the gamma-ray flux UL while keeping the spectral index fixed at $\alpha = 2.3$ and $\ E_{\rm th} = 0.1 \ \mathrm{GeV}$. The results are shown for two different values of $\ E_{\rm cut}$, whereas the solid lines indicate simulations with $ E_{\rm cut} = Z \times 10^{21}\  \mathrm{eV}$ and dashed lines represent $ E_{\rm cut} = Z \times 10^{20.5}\ \mathrm{eV}$. The blue lines indicate a proton-dominated composition, whereas the red lines correspond to an iron-dominated composition. The flux produced by proton sources is significantly higher than iron sources, as protons undergo more efficient interactions that generate secondary gamma rays, while iron nuclei suffer greater deflections and energy losses. The shaded region represents the uncertainty associated with the integral values, computed at a CL 95\% using a chi-square distribution. This figure highlights the dependence of the predicted gamma-ray flux on the primary cosmic-ray composition, as well as the significant impact of the cutoff energy in determining the flux level.

Figure~\ref{fig:1dgamma}-b investigates the influence of the spectral index $\alpha$ of the source spectrum while keeping $\ E_{\rm cut} = Z \times 10^{21}\  \mathrm{eV}$ and the energy threshold at $ E_{\rm th} = 330\  \mathrm{GeV}$.  The simulations consider three values of $\alpha = 2.0$, 2.3, and 2.8. The results show that harder spectra ($\alpha = 2.0$) lead to a higher gamma-ray flux, while softer spectra ($\alpha = 2.8$) result in a lower flux. This behavior is expected since a harder spectrum means that a larger fraction of the UHECR energy is concentrated at higher energies, enhancing the production of secondary gamma rays. Once again, the proton-dominated scenario consistently leads to a higher gamma-ray flux than in the iron-rich case. These results emphasize the strong dependence of multimessenger signals on the spectral characteristics of UHECR sources.

The final revisited analysis, shown in Figure~\ref{fig:1dgamma}-c, examines the impact of varying the gamma-ray detection threshold $ E_{\rm th}$ while keeping $ E_{\rm cut} = Z \times 10^{21}\  \mathrm{eV}$ and $\alpha = 2.3$. The analysis considers three energy thresholds: $100$, $300$ and $500$ $\mathrm{GeV}$. As expected, increasing $ E_{\rm th}$ results in a lower gamma-ray flux, as fewer secondary photons exceed the threshold and contribute to the observed flux. The figure reinforces that the choice of observational parameters can significantly influence the detected flux, making precise energy threshold determinations essential for maximizing the detection of gamma-ray signals from UHECR interactions.

Together, these figures illustrate the dependencies of the expected gamma-ray flux on astrophysical parameters such as the UHECR composition, spectral index, and energy threshold. Proton-rich sources consistently yield higher fluxes than iron-rich sources, a trend that persists across all examined variations. A higher cutoff energy results in a stronger flux, while a softer spectral index reduces it. The choice of gamma-ray detection thresholds also plays a crucial role in determining the observable flux, demonstrating that instrumental sensitivity is a key factor in successful multimessenger studies. Future observations with high-energy gamma-ray telescopes, such as the CTAO \citep{2019scta.book.....C}, will be instrumental in validating these predictions and further constraining the nature of UHECR sources.

\subsection{Upper Limits on the integral Neutrino flux for Individual Sources.}\label{2.3}
In this paper, an additional objective was to establish a correlation between the neutrino flux measured and the method used to calculate the UL on cosmic-ray luminosity. Similar to gamma rays, neutrinos serve as messenger particles that interact minimally with the intergalactic medium during their propagation. As a result, it can be inferred that the propagation of UHECR particles from a source generates two distinct secondary fluxes: gamma rays, as described by Equation \ref{eq:gamma}, and a secondary neutrino flux, which can be expressed as follows:
\begin{equation} \label{integralneutrino}
    I_{\nu}(E_{\nu}) = \frac{L_{\mathrm{CR}}}{4 \pi D^{2} (1 + z) \langle E \rangle_{0}} K_{\nu} P_{\nu}(E_{\nu}).
\end{equation}
In this equation, ($K_{\nu}$) represents the number of neutrinos produced per cosmic-ray at the source, while \( P_{\nu}(E_{\nu}) \) describes the energy distribution of neutrinos reaching Earth. Here, \(  E_{\nu} \) denotes the neutrino energy, \( D \) is the comoving distance to the source, and \(  z \) is its redshift. Once the maximum UHECR luminosity of a source is determined via gamma rays, the corresponding upper limit on the neutrino flux can be calculated. This formulation provides insight into the emission of neutrinos and gamma rays as secondary particles generated during cosmic-ray propagation.

The method enables us to determine ULs on the total UHECR luminosity for individual sources. These results will be utilized to derive ULs on the integral neutrino flux for each source, providing critical constraints on their potential neutrino emissions.
\begin{equation} \label{ulneutrino}
    I_{\nu}^{\mathrm{UL}}(> E_{\nu}^{\mathrm{th}}) = L_{\mathrm{CR}}^{\mathrm{UL}} \frac{K_{\nu}\int_{E_{\nu}^{\mathrm{th}}}^{\infty} dE_{\nu} P_{\nu}(E_{\nu})}{4\pi D^{2} (1 + z)\langle E \rangle_{0}}.
\end{equation}

The equation referenced in Equation \ref{ulneutrino} is derived by reformulating Equation \ref{eq:LCR}, incorporating the integral neutrino flux obtained from the simulation framework. These equations are applied to derive results for the NGC 1068 source. Recent observations from the IceCube Observatory \citep{doi:10.1126/science.abg3395} indicate the presence of potential high-energy neutrino emissions originating from this galaxy. Several studies suggest that lepto-hadronic acceleration models, which consider the presence of an AGN at the galaxy's core, could explain this emission. The interaction of high-energy protons within the AGN environment may lead to the production of high-energy neutrinos, specifically linked to this source \citep{anchordoqui2021highenergyneutrinosngc1068,Eichmann_2022}. The current analysis assumes that hadronic interactions allow the most energetic cosmic-ray particles to escape their source and subsequently generate secondary particles \citep{km3net_2025}. These secondary neutrinos propagate through space until reaching Earth, contributing to the secondary flux described by Equation \ref{integralneutrino}. Given that no direct measurement of this flux has yet been performed, we establish ULs on the integral neutrino flux, as defined in Equation \ref{ulneutrino}.

\begin{figure}[htb]
\centering
\includegraphics[width=0.47\textwidth]{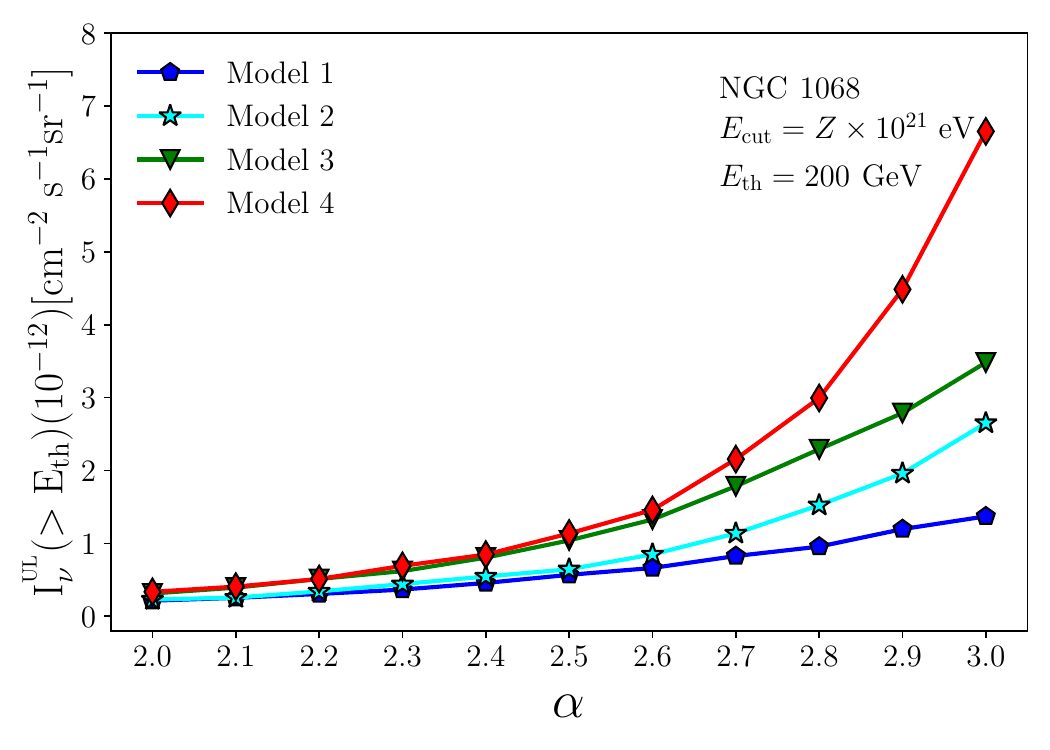}
\caption{UL on the integral neutrino flux ($I\rm_{\nu}^{\mathrm{UL}}$) as a function of the spectral index at the distance of NGC 1068. The maximum energy of the injected particles is set to $E_{\mathrm{cut}} = Z \times 10^{21} \ \mathrm{eV}$, while the energy threshold used in the integral calculation is $E\rm_{\mathrm{th}} = 200 \ \mathrm{GeV}$, as defined by MAGIC~\citep{2019ApJ...883..135A}.}
\label{fig:ULNeutrinos}
\end{figure}

Figure~\ref{fig:ULNeutrinos} presents the ULs on the integral neutrino flux, obtained using the method described in this Section, applying the simulation framework at the distance of NGC 1068 and considering the mass compositions outlined in Table~\ref{tab:1}. The results indicate that the proton fraction in the composition significantly influences the derived ULs on the neutrino flux, suggesting that an increase in the proton content leads to a corresponding increase in the ULs. This observation may point to the presence of hadronic interactions as the dominant mechanism for high-energy neutrino production. Numerous studies have examined the emission of high-energy neutrinos from NGC 1068, providing strong evidence that it is a significant source of high-energy neutrinos, most likely generated through hadronic interactions in dense and obscured regions near the AGN's core~\citep{2023JCAP...06..004C, 2023ApJ...956....8F, 2024ApJ...972...44D}. In particular, \cite{2023JCAP...06..004C} demonstrates that dark matter-neutrino interactions do not significantly attenuate the neutrino flux, ensuring that our calculated ULs on the integral neutrino flux remain consistent with observed astrophysical constraints and multimessenger analyses.

\subsection{Neutrino Integral Flux study}

\begin{figure}[htbp!]
    \centering
    \begin{minipage}{0.47\textwidth} 
        \centering
        \includegraphics[width=\textwidth]{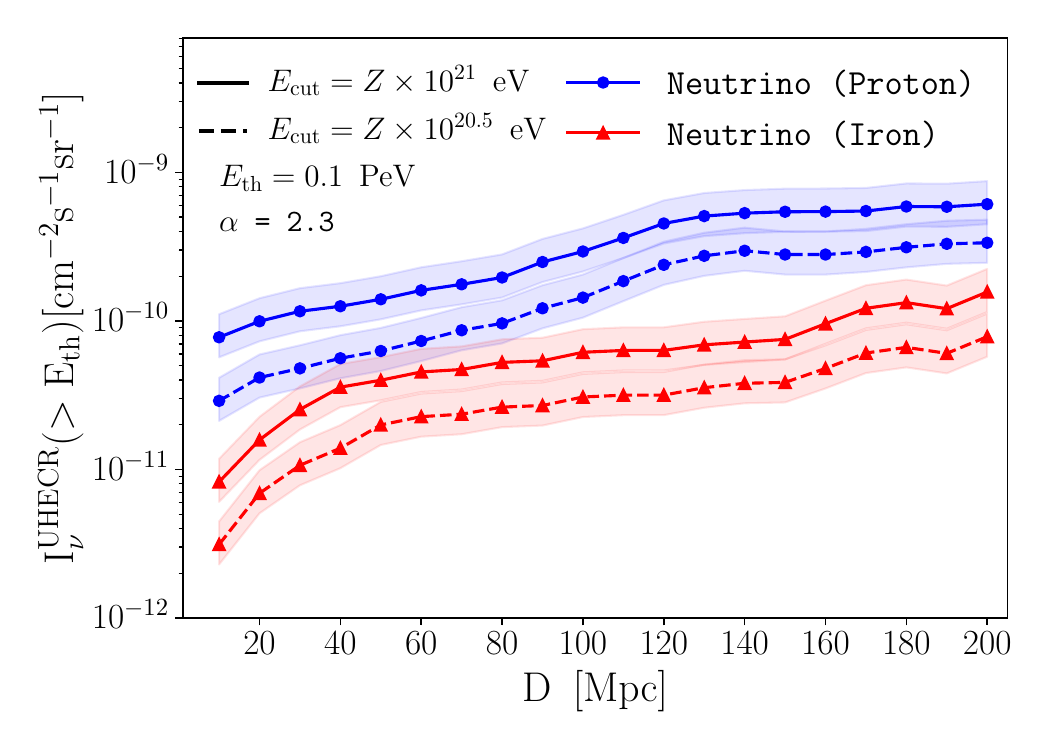} 
        (a) $ E_{\rm cut}$ study.
    \end{minipage}
    \begin{minipage}{0.47\textwidth}
        \centering
        \includegraphics[width=\textwidth]{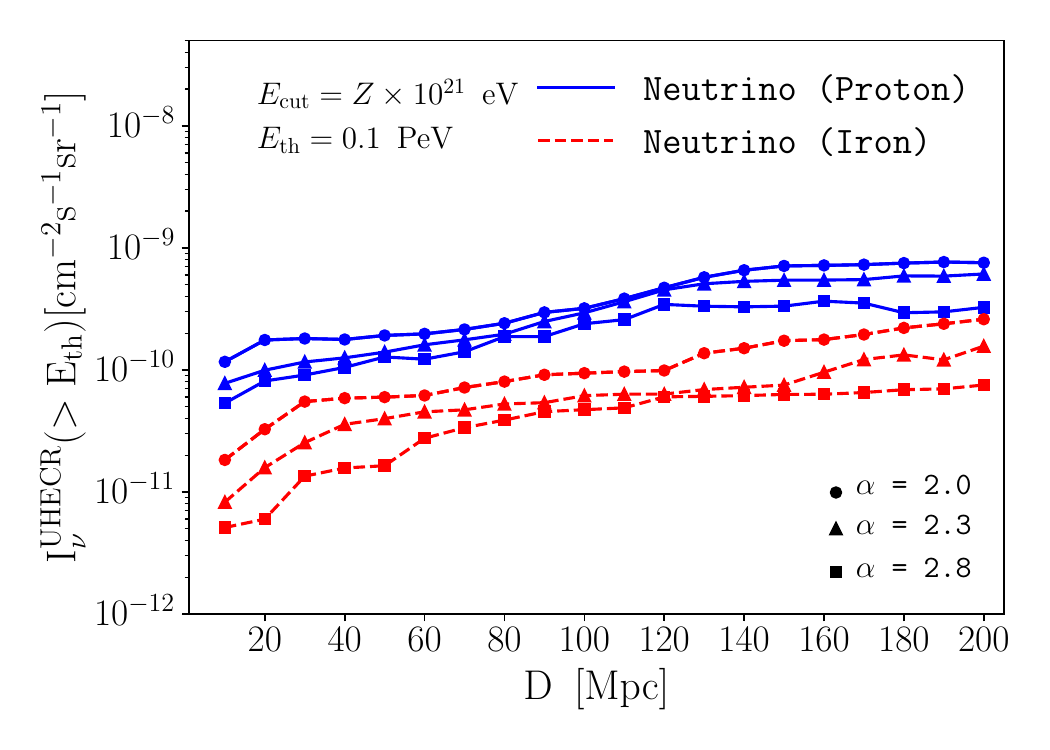}
        (b) $\alpha$ study
    \end{minipage}
    \hfill
    \begin{minipage}{0.47\textwidth}
        \centering
        \includegraphics[width=\textwidth]{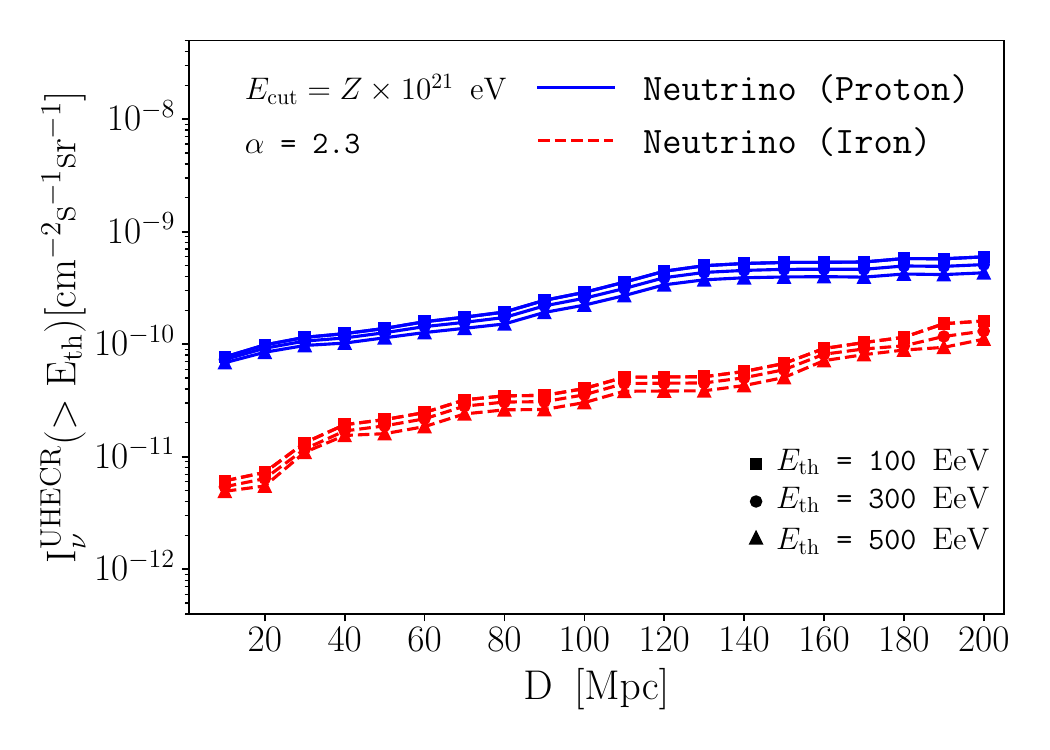}
        (c) $E\rm _{\mathrm{th}}$ study
    \end{minipage}
    \caption{Upper limits (ULs) on the neutrino flux,  
$I^{\mathrm{UHECR}}_{\nu}(> E_{\mathrm{th}})$, as a function of the source distance,  
calculated using the UL on the flux observed by the Pierre Auger Observatory  
at the 95\% confidence level (CL) for a given energy threshold $ E_{\mathrm{th}}$. 
(a) Dependence on the cutoff energy $E_{\mathrm{cut}}$,  
with fixed parameters $\alpha = 2.3$ and $E_{\mathrm{th}} = 0.1 \ \mathrm{PeV}$.  
(b) Dependence on the spectral index $\alpha$,  
for fixed parameters $E_{\mathrm{cut}} = Z \times 10^{21} \ \mathrm{eV}$  
and $E_{\mathrm{th}} = 0.1 \ \mathrm{PeV}$.  
(c) Dependence on the energy threshold $E_{\mathrm{th}}$,  
for fixed parameters $E_{\mathrm{cut}} = Z \times 10^{21} \ \mathrm{eV}$  
and $\alpha = 2.3$.}  
\label{fig:1dneutrino}
\end{figure}

Figure~\ref{fig:1dneutrino}-a illustrates the integral neutrino flux, calculated using Equation~\ref{integralneutrino}. In this section, we present $ I^{\rm UHECR}_{\nu}(>$ $E_{\rm th})$, derived from the UL flux at 95\% CL from the PAO. The analysis in the figure~\ref{fig:1dneutrino}-a considers two different values of $ E_{\rm cut}$ for two compositions. The blue lines correspond to a proton composition, while the red lines represent an iron composition for the injected source particles. Solid lines indicate simulations with $ E_{\rm cut} = Z \times 10^{21} \ \mathrm{eV}$, while dashed lines correspond to $ E_{\rm cut} = Z \times 10^{20.5} \ \mathrm{eV}$. The shaded region represents the uncertainty in the integral values, estimated at 95\% CL using a chi-squared distribution.

Figure~\ref{fig:1dneutrino}-b presents the dependence of $ I^{\rm UHECR}_{\nu}(> E_{\rm th})$ on the spectral index $\alpha$ of the source spectrum across different simulated distances. The spectral indices considered are $\alpha = 2.0, \ 2.3, \ 2.8$. This figure highlights how variations in the source spectrum influence the expected neutrino flux for both proton and iron compositions.

The final analysis, shown in Figure~\ref{fig:1dneutrino}-c, examines the dependence of $ I^{\rm UHECR}_{\nu}(>$ $E_{\rm th})$ on different energy thresholds $ E_{\rm th}$. This study shows that the choice $ E_{\rm th}$ can change the predicted neutrino flux by up to a factor of two, thereby directly affecting the detectability of these signals. The results have important implications for high-energy neutrino observatories such as IceCube~\citep{doi:10.1126/science.abg3395} and KM3NeT~\citep{km3net_2025}, which aim to measure neutrino fluxes from astrophysical sources.

In general, Figures~\ref{fig:1dgamma} and~\ref{fig:1dneutrino} exhibit similar trends, particularly in relation to the influence of composition, spectral index, and energy threshold on secondary fluxes. Both gamma-ray and neutrino fluxes show a higher intensity for proton compositions compared to iron, with a clear dependence on $ E_{\rm cut}$. However, an important distinction is that the gamma-ray fluxes, Figure~\ref{fig:1dgamma}, are significantly affected by electromagnetic interactions, including pair production and inverse Compton scattering, which alter their propagation and reduce their flux at high energies. In contrast, neutrinos, Figure~\ref{fig:1dneutrino}, interact minimally with the intergalactic medium, preserving their flux over long distances. This difference makes neutrinos a more direct probe of the cosmic-ray acceleration mechanisms at the source, while gamma rays carry additional information about propagation effects. Additionally, the energy threshold ($ E_{\rm th}$) has a greater impact on gamma-ray flux predictions than on neutrinos, as gamma-ray fluxes are more sensitive to absorption processes. This comparison highlights the complementarity between gamma-ray and neutrino observations in multimessenger studies of UHECRs sources.

\subsection{Extragalactic Magnetic Field Simulation}\label{2.5}

In order to investigate new scenarios, it is crucial to assess whether meaningful results can be obtained using an EGMF simulation framework~\citep{10.1063/1.4789861, Alves_Batista_2022}. In this section, we aim to determine the maximum EGMF strength that can be applied while ensuring the model remains valid and conservative, preserving the integrity of cosmic-ray propagation predictions.

For this study, we used a turbulent magnetic field spectrum following a power-law distribution across different turbulence scales, defined by $L\rm_{\mathrm{max}}$ and $L\rm_{\mathrm{min}}$, which represent the largest and smallest coherence lengths, respectively. This simulation operates in a four-dimensional framework, requiring the specification of key parameters and the definition of the simulation geometry. For the coherence length values, we set $L\rm_{\mathrm{min}} = 60$ kpc and $L\rm_{\mathrm{max}} = 365$ kpc, ensuring an average coherence length equivalent to the observer's size. The spectral index of the Kolmogorov turbulence spectrum is fixed at $\rm k = 5/3$, which is a commonly adopted value in astrophysical simulations \citep{10.1063/1.4789861, Das_2020, Das_2022}. To assess the impact of the EGMF on particle propagation, we tested different values of its root mean square (RMS) strength in the nG scale. We configured our 3D simulations using a \textit{ObserverSurface} module. In this setup, the observer is a spherical surface (detector) of radius $\mathrm{Obs}_{\mathrm{size}}$,  centered and positioned at each propagation distance under study. We set $\mathrm{Obs}_{\mathrm{size}} = 0.1\;\rm Mpc$. Additionally, we analyze the source spectrum to evaluate its influence on the production of secondary particles~\citep{10.1063/1.4789861, Alves_Batista_2022}. 

\begin{figure}[htbp!]
\centering
\includegraphics[width=0.47\textwidth]{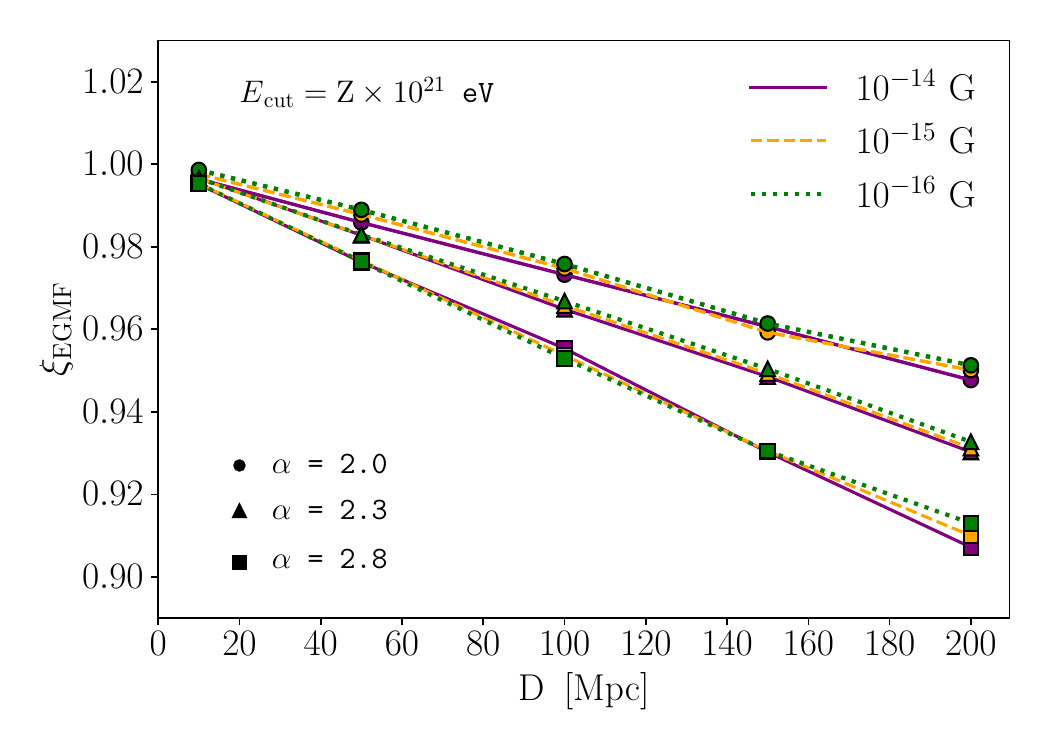}
\caption{Fraction of injected protons $\xi_{\mathrm{EGMF}}$, as a function of distance $D$, with a cutoff energy of $E_{\mathrm{cut}} = Z \times 10^{21} \ \mathrm{eV}$ that successfully reach the observer, shown as a function of source distance for different values of the spectral index $\alpha$.}
\label{fig:fraction}
\end{figure}

The objective is to establish the maximum RMS field strength of the EGMF that can be applied while maintaining a conservative approach. A key criterion for ensuring the validity of the model is that at least 90\% of the particles injected from the source reach the observer. In this study, we adopt a maximum EGMF strength of \(10^{-14}\) G, considering a source distance of 200 Mpc. Under these conditions, the fraction of particles that successfully reach the observer is 91\%, as shown in Figure~\ref{fig:fraction}. For shorter distances, this fraction increases, allowing for the use of stronger EGMF strengths in further analyses. While the typical estimated value of the EGMF strength is around 1.0 nG~\citep{10.1093/mnras/stx3354, 10.1093/mnras/stab3495}, several studies suggest significantly lower bounds values, ranging from approximately \(10^{-14}\) to \(10^{-17}\) G. These references show that TeV $\gamma$-rays measurements of blazars absorbed by EBL photons produce an electromagnetic cascade via secondary electrons and positrons, on which the observed flux depends on the EGMF parameters, making it a powerful tool for constraining field strength values on EGMF away from clusters and filaments on large-scale structure.~\citep[see]{Bray_2018, 10.1093/mnras/stac2509, nature2024}. Recent publications have combined data from H.E.S.S and Fermi-LAT and performed the first combined likelihood fit for multiple sources, providing more robust constraints~\citep{HESS:2023afw}. In addition, the MAGIC Collaborations has investigated how the intrinsic variability of the source in the VHE band impacts the EGMF limits, adding a crucial time-dependent dimension to these mensurements~\citep{MAGIC:2022piy}. Figure \ref{fig:fraction} illustrates the fraction of injected protons, denoted as $\xi_{\mathrm{EGMF}}$, that successfully reach the observer as a function of source distance for different values of the EGMF strength. The analysis considers a cutoff energy of $ E_{\mathrm{cut}} = Z \times 10^{21} \ \mathrm{eV}$ and evaluates three spectral indices: $\alpha = 2.0, \ 2.3,\  2.8$. The curves correspond to EGMF strengths of $10^{-14} \ \mathrm{G}$ (purple), $10^{-15} \ \mathrm{G}$ (orange), and $10^{-16} \ \mathrm{G}$ (green), which represent different levels of magnetic field influence on cosmic-ray propagation. The results show that as the source distance increases, the fraction of particles reaching the observer decreases due to greater deflections caused by the magnetic field. For closer sources, a higher percentage of protons arrives at the observer, allowing for stronger EGMF values to be considered in the simulations. These results clarify the relationship between magnetic deflections and cosmic-ray propagation, constraining EGMF strengths to maintain model validity while ensuring a sufficient number of cosmic rays reach the observer.

This choice of a maximum EGMF strength of \(1 \times 10^{-14} \ \mathrm{G}\) is not arbitrary, but rather reflects a necessary condition to preserve the physical validity of the method. Stronger magnetic fields would result in angular deflections so large that the arrival directions of UHECRs and their secondary gamma rays would no longer correlate with the source direction, violating the underlying assumption of directional association. Under such conditions, Equation~\ref{eq:LCR} becomes inapplicable, as the gamma-ray flux could no longer be attributed to a specific point source. Therefore, our results are presented as conditional upper limits, valid only within the regime where magnetic deflections remain moderate and at least 90\% of the injected UHECRs reach the observer. This ensures that the association between source and secondary flux remains meaningful, while maintaining consistency across the different propagation scenarios studied.

The fractions presented are calculated based on the deflections of UHECRs that reach the observer, denoted as $\mathrm{Obs}_{\mathrm{size}}$. These deflections are estimated by analyzing the initial and final momentum vector directions obtained from the simulation. The choice of applying lower values for the EGMF strength is guided by the extent of UHECR deflections, ensuring a conservative approach in propagation modeling. The RMS deflection of cosmic-ray trajectories, while propagating over a distance $D$ in a randomly turbulent magnetic field, is given by the expression \citep{Dermer_2009}:
\begin{equation}\label{deflection}
    \Phi_{\mathrm{RMS}} \approx 4^{\circ}\frac{10^{3}\ \mathrm{EeV}}{E/Z}\frac{B_\mathrm{{RMS}}}{10^{-9} \mathrm{G}}\sqrt{\frac{D}{\mathrm{100\ \mathrm{Mpc}}}}\sqrt{\frac{L_\mathrm{c}}{1\ \mathrm{Mpc}}},
\end{equation}
where $L_{\mathrm{c}}$ represents the coherence length, set to 0.1 Mpc in this study. Applying this equation, we consider a proton with a maximum energy of $ E_{\mathrm{max}} = 10^3$ EeV propagating over a distance of 200 Mpc through an EGMF with an RMS strength of $\rm B_{\mathrm{RMS}} = 10^{-14}$ G. Under these conditions, the estimated deflection is approximately $0.0178^{\circ}$. The simulation environment for the specific case involving the EGMFs includes a special configuration for particle emission, featuring a cone-shaped setup directed toward the observer. The cone's opening angle is determined by the values obtained from equation~\ref{deflection} for each simulation configuration. This approach provides a conservative estimate of the impact of magnetic deflections by avoiding additional angular dispersion, which would otherwise reduce the fraction of arriving particles. Consequently, the model prevents an overestimation of the secondary gamma-ray and neutrino fluxes that could result from large-angle deflections.

\subsection{EGMF $\gamma$-ray study}\label{2.6}

\begin{figure*}[ht]
    \centering
    \begin{minipage}{0.49\textwidth}
        \centering
        \includegraphics[width=\textwidth]{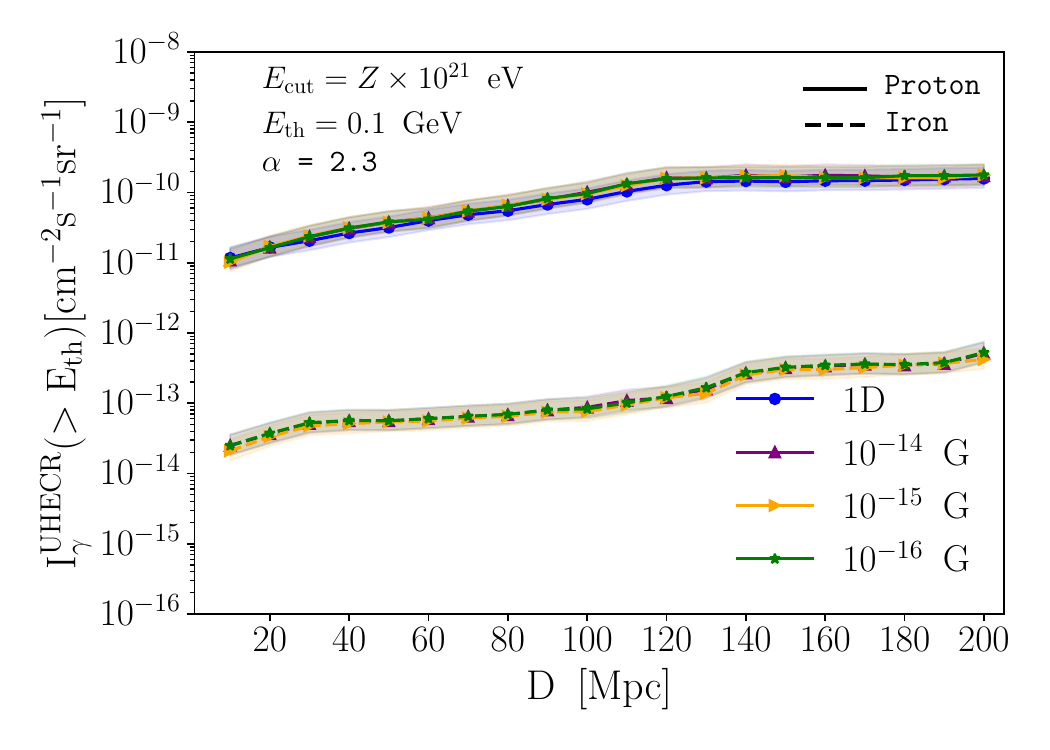}
        \par\vspace{5pt}
        (a) EGMF fixed parameters.
    \end{minipage}
    \begin{minipage}{0.49\textwidth} 
        \centering
        \includegraphics[width=\textwidth]{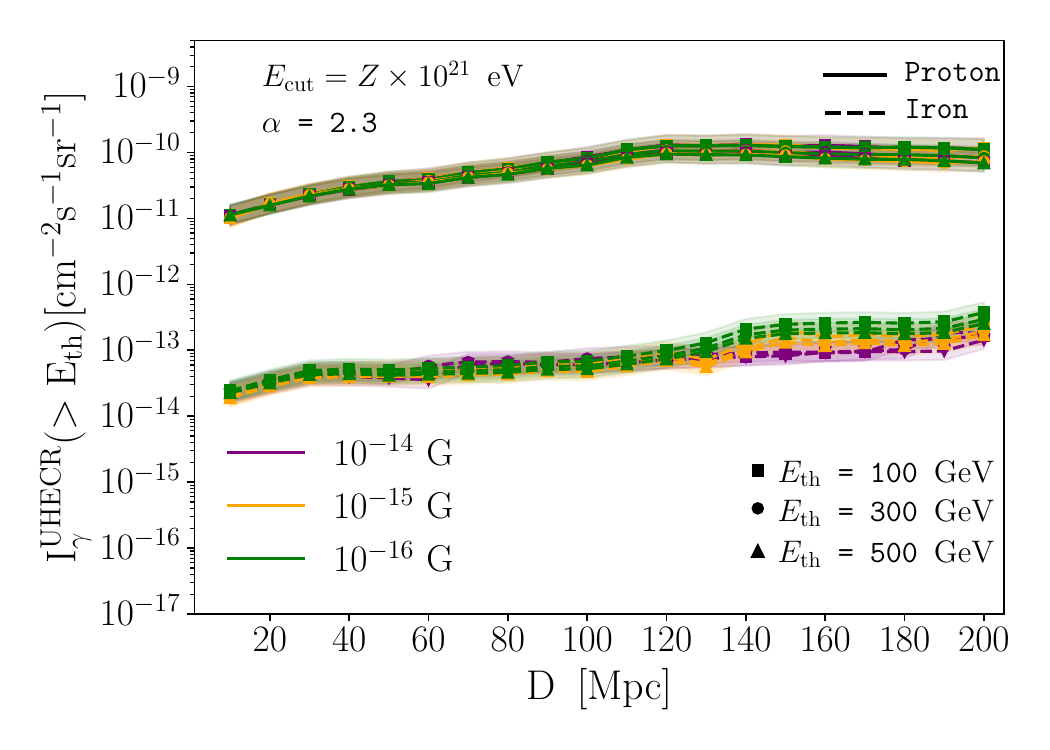} 
        \par\vspace{5pt} 
        (b) $ E_{\rm th}$ study
    \end{minipage}
    \hfill 
    \hfill

    \begin{minipage}{0.49\textwidth}
        \centering
        \includegraphics[width=\textwidth]{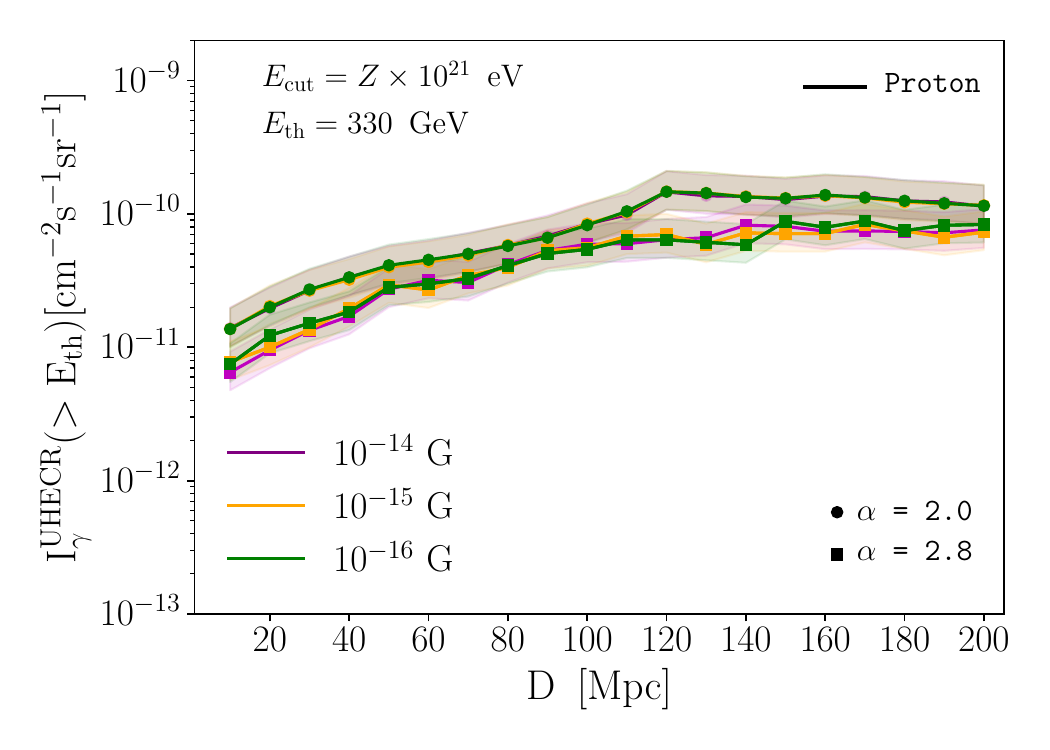}
        \par\vspace{5pt}
        (c) Proton - ${\alpha}$ study
    \end{minipage}
    \begin{minipage}{0.49\textwidth}
        \centering
        \includegraphics[width=\textwidth]{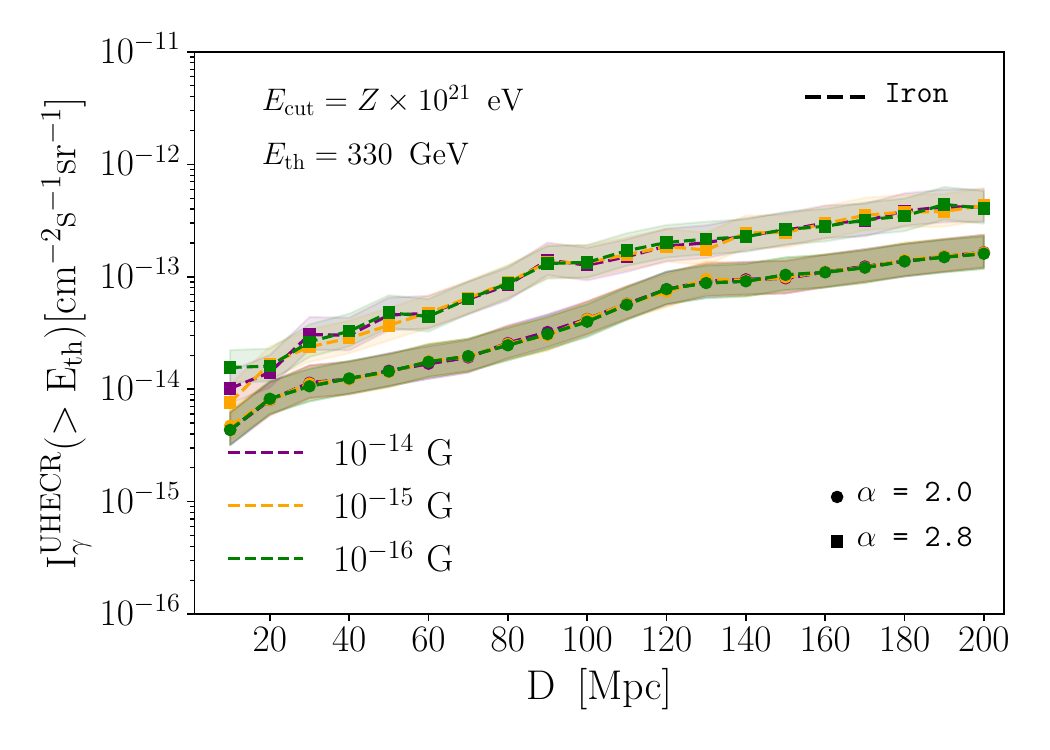}
        \par\vspace{5pt}
        (d) Iron - ${\alpha}$ study
    \end{minipage}
    \caption{The plots illustrate the integral gamma-ray flux $\ I^{\mathrm{UHECR}}_{\gamma}$ as a function of the source distance, calculated using the UL on the flux observed by the PAO at  95\% CL within the EGMF simulation framework. (a) Presents $ I^{\mathrm{UHECR}}_{\gamma}$ for fixed parameters, with $E\rm _{\mathrm{cut}}$ and spectral index $\alpha = 2.3$, and an energy threshold of $E\rm_{\mathrm{th}} = 0.1 \ \mathrm{GeV}$. For comparative purposes, the solid blue line represents $I\rm^{\mathrm{UHECR}}_{\gamma}$ obtained using the 1D simulation framework. (b) Explores the dependence of $I\rm^{\mathrm{UHECR}}_{\gamma}(> E\rm_{\mathrm{th}})$ on the energy threshold, keeping fixed parameters of $E\rm_{\mathrm{cut}} = Z \times 10^{21} \ \mathrm{eV}$ and $\alpha = 2.3$. (c) and (d) Analyze the effect of varying the spectral index $\alpha$ on $I^{\mathrm{UHECR}}_{\gamma}(>E_{\mathrm{th}})$. Both subfigures use fixed parameters of $E\rm_{\mathrm{cut}} = Z \times 10^{21} \ \mathrm{eV}$ and an energy threshold of $E\rm_{\mathrm{th}} = 330 \ \mathrm{GeV}$. Figure (c) corresponds to the proton composition, while Figure (d) represents the iron composition. The shaded regions in all plots indicate the uncertainty associated with the integral values, estimated at 95\% CL using a chi-squared distribution.}
    \label{fig:egmf}
\end{figure*}

Figure~\ref{fig:egmf} presents the impact of the EGMF on the gamma-ray flux predictions, specifically the integral flux $ I^{\mathrm{UHECR}}_{\gamma}(>$ $E_{\mathrm{th}})$, while ensuring that the chosen EGMF values maintain the conservative method by allowing at least 90\% of injected UHECRs to reach the observer. These results are based on the upper UL flux at a 95\% CL from the PAO. The study systematically analyzes the effects of different EGMF strengths, spectral indices, and energy thresholds on gamma-ray flux predictions. Subfigure~\ref{fig:egmf}-a presents $ I^{\mathrm{UHECR}}_{\gamma}(>$$ E_{\mathrm{th}})$ for different EGMF strengths, assuming fixed spectral parameters of $\alpha = 2.3$, $ E_{\mathrm{cut}} = Z \times 10^{21} \ \mathrm{eV}$, and $ E_{\mathrm{th}} = 0.1 \ \mathrm{GeV}$. The colors represent different field strengths: $B_{\mathrm{RMS}} = 1 \times 10^{-14} \ \mathrm{G}$ (purple line), $B_{\mathrm{RMS}} = 1 \times 10^{-15} \ \mathrm{G}$ (orange line), and $B_{\mathrm{RMS}} = 1 \times 10^{-16} \ \mathrm{G}$ (green line). The solid lines indicate a proton composition, while the dashed lines correspond to an iron composition for the injected particles.  

Subfigure~\ref{fig:egmf}-b shows the variation of $I^{\mathrm{UHECR}}_{\gamma}(> $$ E_{\mathrm{th}})$ with different gamma-ray energy thresholds while keeping $ E_{\mathrm{cut}} = Z \times 10^{21} \ \mathrm{eV}$ and $\alpha = 2.3$ constant. The analysis highlights the strong impact of $ E_{\mathrm{th}}$ on gamma-ray flux measurements, with implications for observatories such as \citep{RC, refId0, 2019ApJ...883..135A}.

Subfigures~\ref{fig:egmf}-c and~\ref{fig:egmf}-d examine the influence of different values for the spectral index $\alpha$ on the gamma-ray flux. The spectral indices considered are $\alpha = 2.0$ and $\alpha = 2.8$, with fixed parameters $ E_{\mathrm{cut}} = Z \times 10^{21} \ \mathrm{eV}$ and $ E_{\mathrm{th}} = 330 \ \mathrm{GeV}$. The results illustrate the distinct dependence of $ I^{\mathrm{UHECR}}_{\gamma}(> $$ E_{\mathrm{th}})$ on spectral index variations, emphasizing differences between proton and iron compositions. The shaded areas across all subfigures represent the uncertainty in integral values, estimated at 95\% CL using a chi-squared distribution. The agreement between the 1D simulation results (Figure~\ref{fig:1dgamma}) and the EGMF-inclusive analysis (Figure~\ref{fig:egmf}) confirms the robustness of the method in estimating UHECR-induced secondary emissions while ensuring its validity for the selected EGMF values under realistic extragalactic conditions.

The methodology developed in this work, while based on a conservative upper limit scenario, can be generalized to account for a broader range of EGMF configurations. Specifically, the derived UL on UHECR luminosity can be modified by introducing an EGMF-dependent efficiency factor, $\xi_{\mathrm{EGMF}}(B, L_c)$, which quantifies the fraction of particles that reach the observer as a function of the magnetic field strength $B$ and coherence length $L_c$. This factor allows users of the method to rescale the luminosity limits according to different assumptions about the EGMF. Figures \ref{fig:fraction} and \ref{fig:egmf} provide a visual reference for how $\xi_{\mathrm{EGMF}}$ varies with source distance and magnetic field strength, enabling practical adjustments to the derived constraints for more optimistic or pessimistic magnetic field scenarios.

\clearpage
\subsection{Galactic Magnetic Field}\label{2.7}

To enhance the robustness of our findings, it is crucial to account the effect on the apparent location of individual sources. This effect arises from the deflection of UHECR as they propagate through both the galactic Magnetic Field (GMF) and the EGMF, an effect that has been previously discussed. Since it can be many orders of magnitude higher in comparison of EGMF strength values, the deflection caused by UHECRs in the GMF can be approximated as

\begin{equation}
    \theta_{\mathrm{def}, \mathrm{MW}} \approx \frac{0.9^{\circ}}{\sin b}\left(\frac{60\ \mathrm{EeV}}{E / Z}\right)\left(\frac{B}{10^{-9}\ \mathrm{G}}\right)\left(\frac{h_{\mathrm{disk}}}{1\ \mathrm{kpc}}\right),
\end{equation}
where b is the galactic latitude of the source and $h_\mathrm{disk}$ is the height of the Galactic disk~\cite{Dermer_2009, Das_2020}.  We include \(\theta_{\mathrm{def,\,MW}}\) in the source weight \(W(\hat{n})\) defined in Section~\ref{section2}, which already incorporates source position, detector exposure, and the factor \(\xi_{\rm EGMF}\). This parameter allows us to make an estimation of the observed event rate for each source by the GMF effect together with the already calculated $\xi_{EGMF}$. The deflection caused by the GMF is estimated using the realistic model JF12, considering the random striated (large-scale) and random turbulent (small-scale) components, calibrated to WMAP7 synchrotron data and extragalactic rotation measures~\cite{Jansson_2012,Unger}. We employ a backtracking technique, a standard method for studying the propagation of charged particles. In this approach, we simulate antiparticles launched from the observer's position back to the edge of the Galaxy. This is computationally equivalent to simulating the trajectories of real particles (with positive charge) arriving at the observer from extragalactic space.

\begin{figure}[h]
\includegraphics[width=0.47\textwidth]{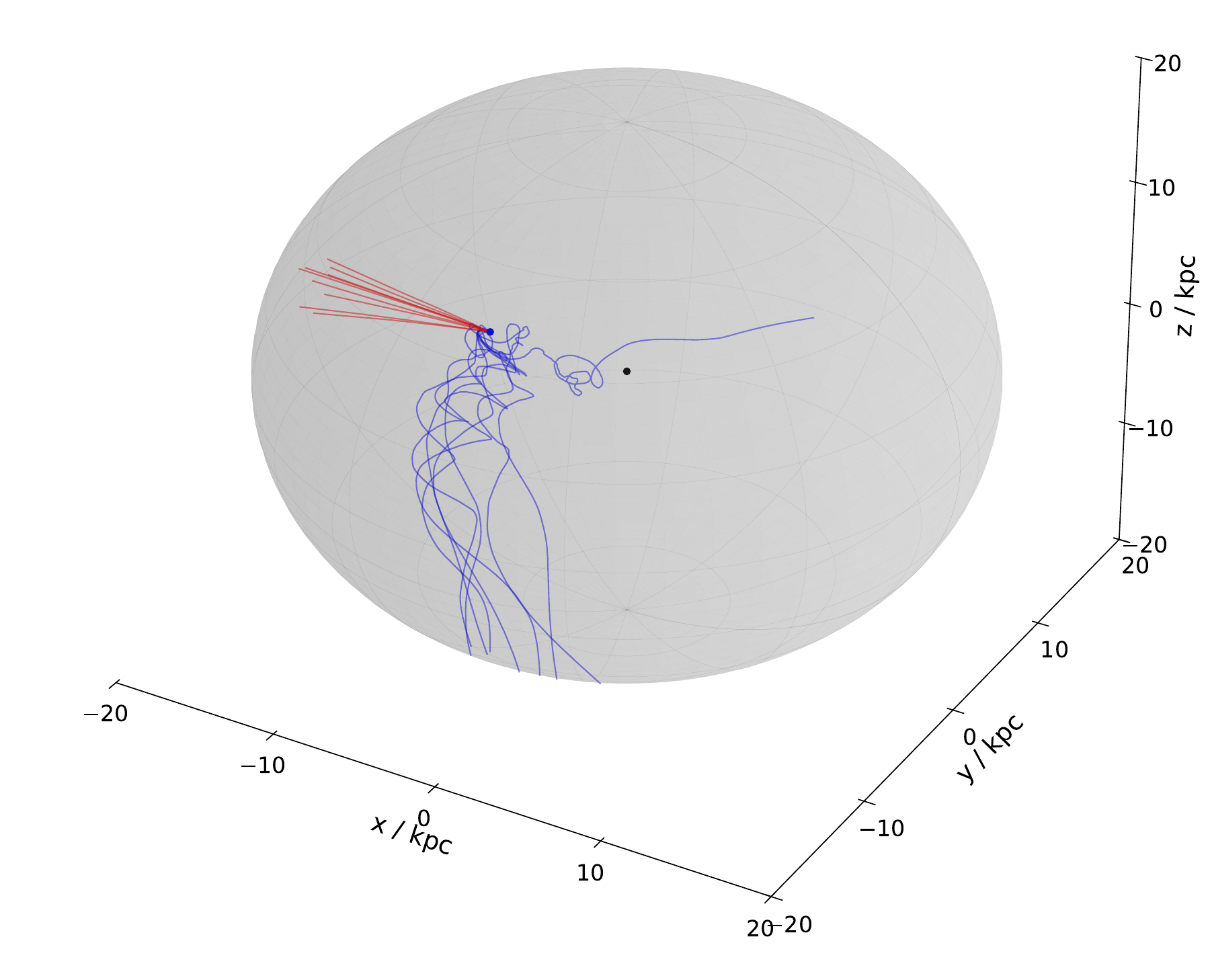}
\caption{An illustration of rigidity-dependent cosmic-ray deflection in the JF12 Galactic magnetic field. Trajectories are back-tracked from an observer at the Solar position ($x=-8.5$ kpc) to a spherical boundary at 20 kpc. The black dot at the sphere’s center marks the Galactic Center. Blue curves show the strongly deflected paths of ten 1 EeV protons ($Z=1$), while red curves show the nearly rectilinear paths of ten 100 EeV protons. The figure shows how the GMF shapes charged-particle propagation within the Galaxy.}
\label{fig:trajectories}
\end{figure}

Figure~\ref{fig:trajectories} provides a visual example of this magnetic deflection, illustrating the dependence on particle rigidity. The simulation shows trajectories for protons with energies of 1 EeV and 100 EeV. These particles are propagated backwards (as antiprotons) from the observer's location within the Galaxy out to a 20 kpc boundary. This demonstrates how lower-energy particles undergo significant deflection, whereas high-energy particles are less affected, supporting a scenario where UHECRs can point back to their sources with less deviation


To account for the effect of the galactic magnetic field, we employ the backtracking technique using the JF12 model to propagate anti-nuclei from the observer's position back to a detection sphere with a radius of 20~kpc. The particle energies are sampled from a power-law spectrum of the form $dN/dE \propto E^{-\gamma}$, with spectral index $\gamma = 2.0$, over the energy range from 1~EeV to 100~EeV. In addition, to model the angular uncertainty in the arrival direction of simulated events, a Gaussian angular deviation with $\sigma = 0.107$ radians is applied. Based on this, we define circular regions with 95\% confidence level around each candidate source position. These angular windows, with radii of $15^\circ$, follow the upper-limit analysis criteria adopted by the Pierre Auger Observatory for UHECR sources~\citep{2019JCAP...10..022A}.

\begin{figure*}[ht]
\centering
\includegraphics[width=0.99\textwidth]{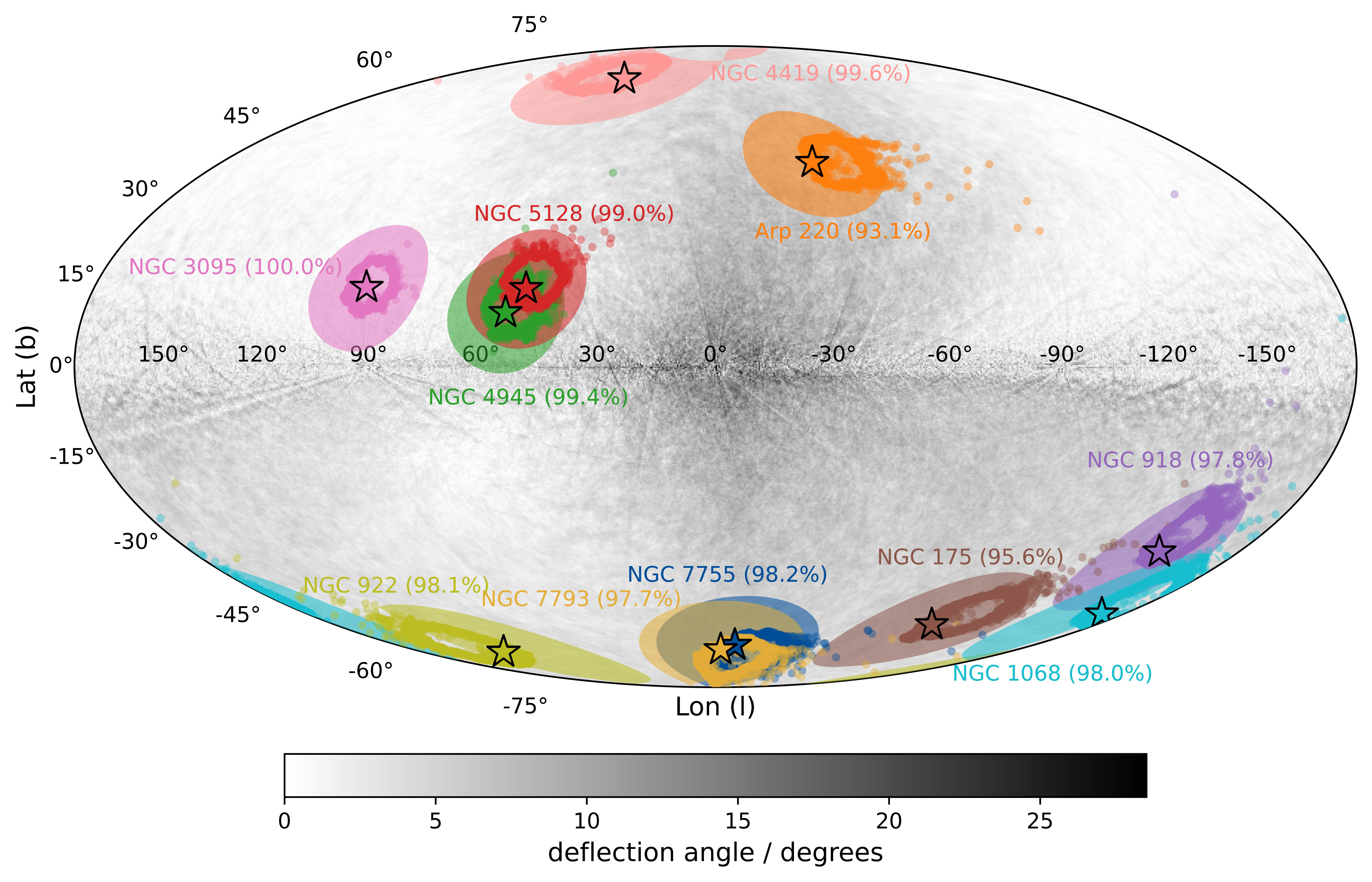}
\caption{Full-sky map in Hammer projection showing the average angular deflection of protons due to the regular and turbulent galactic magnetic field, modeled using the JF12 framework. The map is presented in galactic coordinates, with the Galactic center located at $(\ell, b) = (0^{\circ}, 0^{\circ})$. Star markers indicate the positions of candidate source galaxies, each labeled with its name. The surrounding colored regions represent the 95\% confidence level contours for the backtracked particle arrival directions. The percentage next to each source indicates the fraction of simulated particles that arrive within the corresponding confidence region.
}
\label{fig:sources_background}
\end{figure*}

Figure~\ref{fig:sources_background} displays the background map of expected deflection angles caused by the galactic magnetic field, based on the JF12 model and simulated backtracked trajectories of antiprotons. The particles follow a defined energy spectrum between 1~EeV and 100~EeV, and the color scale represents the average deflection angle across the sky. A strong anisotropy is evident, especially near the Galactic center, where the disk component of the magnetic field is most intense. The large-scale asymmetries arise from the combined effects of the regular, striated, and turbulent components of the GMF~\cite{Jansson_2012}. Colored stars mark the positions of candidate sources. For each source, the shaded region denotes the 95\% CL area around its location, and the small dots show the individual arrival directions of $10^3$ backtracked antiprotons. These arrival patterns clearly illustrate the shadowing effect induced by deflections, forming a spatial probability distribution that characterizes the expected UHECR arrival direction at Earth. The percentages next to each source indicate the fraction of particles, $\xi_{\mathrm{GMF}}$, that arrive within the 95\% CL region despite magnetic deflection. This factor is used to weight the cosmic-ray flux when computing the effective upper limit on cosmic-ray luminosity, $L^{\mathrm{eff}}_{\mathrm{CR}}$, for each source. Starting from Equation~\ref{eq:LCR}, which defines the upper limit $L^{\mathrm{UL}}_{\mathrm{CR}}$, we introduce the correction due to both extragalactic and galactic magnetic fields by including $\xi_{\mathrm{EGMF}}$ and $\xi_{\mathrm{GMF}}$:

\begin{equation}
    L_{\mathrm{UHECR}}^{\mathrm{eff}} = \frac{4 \pi D^{2} (1 + z) \langle E \rangle_{0}}{\int_{E_{\gamma}^{\mathrm{th}}}^{\infty} dE \, P_{\gamma}(E_{\gamma})} I_{\gamma}^{\mathrm{UL}}(> E_{\gamma}^{\mathrm{th}}) \times \xi_{total},
\end{equation}
where $\xi_{total}$ is defined as:
\begin{equation}
    \xi_{\mathrm{total}} = \xi_{\mathrm{EGMF}} \hspace{0.1cm} \times \hspace{0.1cm}\xi_{\mathrm{GMF}}.
\end{equation}
This procedure accounts for the effects of magnetic fields on UHECR propagation by weighting the total cosmic-ray luminosity inferred from cosmogenic gamma rays according to the fraction of injected particles that successfully reach the observer. In doing so, the method maintains a conservative approach, ensuring that only the portion of particles effectively contributing to the observed flux is considered for each individual source. The resulting effective upper limits on CR luminosity, $L_{\mathrm{CR}}^{\mathrm{eff}}$, for the selected sources are presented in Section~\ref{individualUL}.

\section{Source Selection}\label{section3}

This section presents the set of astrophysical sources selected for our analysis. We focus on eleven nearby candidate sources with potential for UHECR acceleration, including starburst galaxies, radio galaxies, and AGNs, chosen for their relevance to high-energy gamma-ray and neutrino observations.

The analysis incorporates the effects of both extragalactic and galactic magnetic fields, as described in Sections~\ref{2.5} and~\ref{2.7}. The following subsections introduce the selected sources and summarize their key properties, along with the corresponding propagation results and deflection efficiency factors used to compute effective upper limits on their CR luminosities. Based on these calculations, we derive the expected gamma-ray flux from UHECRs, $I_{\gamma}^{\mathrm{UHECR}}$, and the associated luminosity limit, $L_{\mathrm{CR}}^{\mathrm{eff}}$, for each source.

\subsection{H.E.S.S. Observatory} 

Based on the ULs of the gamma-ray flux obtained by the H.E.S.S. Observatory~\citep{refId0} for different sources, we apply the method of estimating the UL on the cosmic-ray luminosity from the UL on the gamma-ray flux. This approach is detailed in Table~\ref{tab:2}, which presents the corresponding measurements. Since the observatory did not detect significant gamma-ray flux from these sources, only ULs were derived. A total of nine galaxies with measured ULs were identified in this reference. Additionally, the starburst galaxy Arp 220 was included as an additional source in this study. It is well known for its intense star formation and serves as a prominent example of the nearest Ultra-Luminous Infrared Galaxy (ULIRG) at $z \simeq 0.018$. The UL on the integral gamma-ray flux of Arp 220, obtained by H.E.S.S. at a 99.9\% confidence level (CL), is given by 
$I_{\gamma}^{\mathrm{UL}}(E_{\gamma}^{\mathrm{th}} > 1 \ \mathrm{TeV}) = 1.32 \times 10^{-14} \ \mathrm{cm^{-2} \ s^{-1}}$ \citep{RC}.

\begin{table}[htbp!]
    \centering
    \caption{Host galaxies and associated supernovae (SN) observed by H.E.S.S. The ULs on the flux are calculated assuming a 95\% CL~\cite{refId0}.}
    \label{tab:2}
    \begin{tabular}{cccc}
        \toprule
        \textbf{Host} & \textbf{SN Name} & \textbf{Dist.} & \textbf{$I_{\gamma}^{\mathrm{UL}}(>E_{\gamma}^{\mathrm{th}}) \times 10^{-13}$} \\
        \vspace{0.005cm}& & (Mpc) & ($\mathrm{cm^{-2}s^{-1}}$) \\
        \midrule
        NGC 7755  & SN 2004cx   & $26.0 \pm 5.00$      & 1.9  \\ 
        NGC 7793  & SN 2008bk   & $4.0 \pm 0.40$   & 4.8  \\    
        NGC 3095  & SN 2008bp   & $29.0 \pm 6.00$      & 5.5  \\ 
        NGC 922   & SN 2008ho   & $41.5 \pm 2.90$  & 7.7  \\  
        NGC 175   & SN 2009hf   & $53.9 \pm 3.80$  & 5.3  \\          
        NGC 918   & SN 2009js   & $16.0 \pm 3.00$      & 11.0 \\ 
        NGC 4945  & SN 2011ja   & $5.28 \pm 0.38$ & 5.2  \\
        NGC 4419  & SN 2012cc   & $16.5 \pm 1.10$  & 10.0 \\
        NGC 5128  & SN 2016adj  & $3.8 \pm 0.10$   & 1.7  \\ 
        \bottomrule
        \end{tabular}
\end{table}

\subsection{MAGIC Observatory}

The data from the MAGIC collaboration, which conducted observations searching for VHE gamma rays, were also utilized for the individual source NGC 1068. After 125 hours of observation, an UL on the gamma-ray flux was obtained at 95\% CL with an energy threshold of $ E_{\mathrm{th}} = 200 \ \mathrm{GeV}$. The derived UL is expressed as: $I_{\gamma}^{\mathrm{UL}}(E_{\gamma}^{\mathrm{th}} > 200 \ \mathrm{GeV})$ = 5.1 $\times$ 10$^{-13}$ (cm$^{2}$ s)$^{-1}$~\citep{2019ApJ...883..135A}.

\subsection{NGC 1068}

NGC 1068 is a nearby type-2 Seyfert galaxy that has been identified as a neutrino source with a CL of 2.9$\sigma$~\citep{murase2019multimessenger, PhysRevLett.124.051103}. In 2022, this significance increased to a global level of 4.2$\sigma$, reinforcing its classification as a potential astrophysical neutrino emitter. The observed event count deviates significantly from the expected background, leading the IceCube Observatory to determine a neutrino flux from this direction
$\phi_{\nu} \approx 3 \times 10^{-8} \left(\frac{E_{\nu}}{\mathrm{TeV}}\right)^{-3.2} \ (\mathrm{GeV \ cm^{-2} \ s^{-1}})$ \citep{doi:10.1126/science.abg3395}.

Since no significant gamma-ray flux has been observed from NGC 1068, we compared the UL on the gamma-ray flux reported by the MAGIC Collaboration with the theoretical flux, $I_{\gamma}^{\mathrm{UHECR}}$, predicted by our models and check the necessary condition criterion detailed in Section~\ref{section2}. Therefore considered an appropriate candidate, NGC 1068 was incorporated into our final analysis about mixed composition in the VHE range. From UL on the cosmic-ray luminosity, we calculated the UL on the integral neutrino flux on the basis of our model (see Section~\ref{2.3}). The minimal interaction of neutrinos with the intergalactic medium during propagation makes them a crucial probe of cosmic-ray acceleration sites. Establishing this correlation provides valuable insights into the nature of cosmic-ray luminosity and its connection to high-energy neutrino production.


\textcolor{black}{Figure~\ref{upperlimits} is central to our limit-setting procedure, as it determines whether we can extract a meaningful bound for each source. For every assumed UHECR composition (proton, iron, and the mixed case discussed in Section~\ref{section2}), we compute the cascade $\gamma$-ray integral flux $I_{\gamma}^{\mathrm{UHECR}}(>E_{\gamma}^{\mathrm{th}})$ after propagation through the CMB and EBL. Because primary VHE photons are attenuated by pair production and their energy is reprocessed to lower energies via electromagnetic cascades, part of the emission is shifted downward in energy but still contributes to the integral flux above $E_{\gamma}^{\mathrm{th}}$. This effect ensures that the total integral flux is conserved, which validates our method of directly comparing the total predicted flux against the observational limits. A meaningful bound can be set only when the unscaled model lies above the observational UL, as this is the condition required for the data to be constraining. Specifically, whenever
\[
I_{\gamma}^{\mathrm{UHECR}}(>E_{\gamma}^{\mathrm{th}}) \;>\; I_{\gamma}^{\mathrm{UL}}(>E_{\gamma}^{\mathrm{th}}),
\]
we uniformly rescale the source luminosity so that
\[
I_{\gamma}^{\mathrm{UHECR}}(>E_{\gamma}^{\mathrm{th}}) \;=\; I_{\gamma}^{\mathrm{UL}}(>E_{\gamma}^{\mathrm{th}}),
\]
and take the corresponding luminosity value as the effective upper limit $L_{\mathrm{CR}}^{\mathrm{eff}}$ (hereafter $L_{\mathrm{CR}}^{\mathrm{UL}}$; see Eq.~\ref{eq:LCR}). If this inequality is not satisfied, the observational data do not provide a meaningful constraint for the assumed composition. A source is retained for the final derivation of its luminosity limit only if its observed $\gamma$-ray upper limit lies below the theoretical prediction corresponding to its assumed UHECR composition, as detailed in Section~\ref{section2}.}

Consequently, we have successfully derived the ULs for all the sources presented in Figure~\ref{upperlimits}. The ULs on the integral neutrino flux, denoted as $I\rm^{\mathrm{UHECR}}_{\nu}$, are also shown. The integral values of the neutrino flux are consistent with the results reported by the PAO ~\cite{Aab_2019}, where a single-flavor integrated limit at a 90\% CL is $4.4 \times 10^{-9} \ \mathrm{GeV \ cm^{-2} \ s^{-1} \ sr^{-1}}$ for a diffuse neutrino source flux. Additionally, the integral values of the neutrino flux are significantly higher than those of the integral gamma-ray flux at the same distances. This discrepancy arises because gamma rays interacting extensively with background radiation, lead to energy losses and attenuation at high energies. In contrast, neutrinos escape without significant interactions, preserving their original flux. This effect highlights the fundamental differences in the physical processes that govern the production and propagation of these particles in the universe. The integrated neutrino flux accounts for the total energy transferred in hadronic interactions, whereas gamma rays suffer energy losses due to absorption and scattering along their path~\citep{vanVliet:2017obm, RBatista_2019, galaxies12060077}.

\hypertarget{upperlimits}{\begin{figure*}[ht]
\centering
\includegraphics[width=0.99\textwidth]{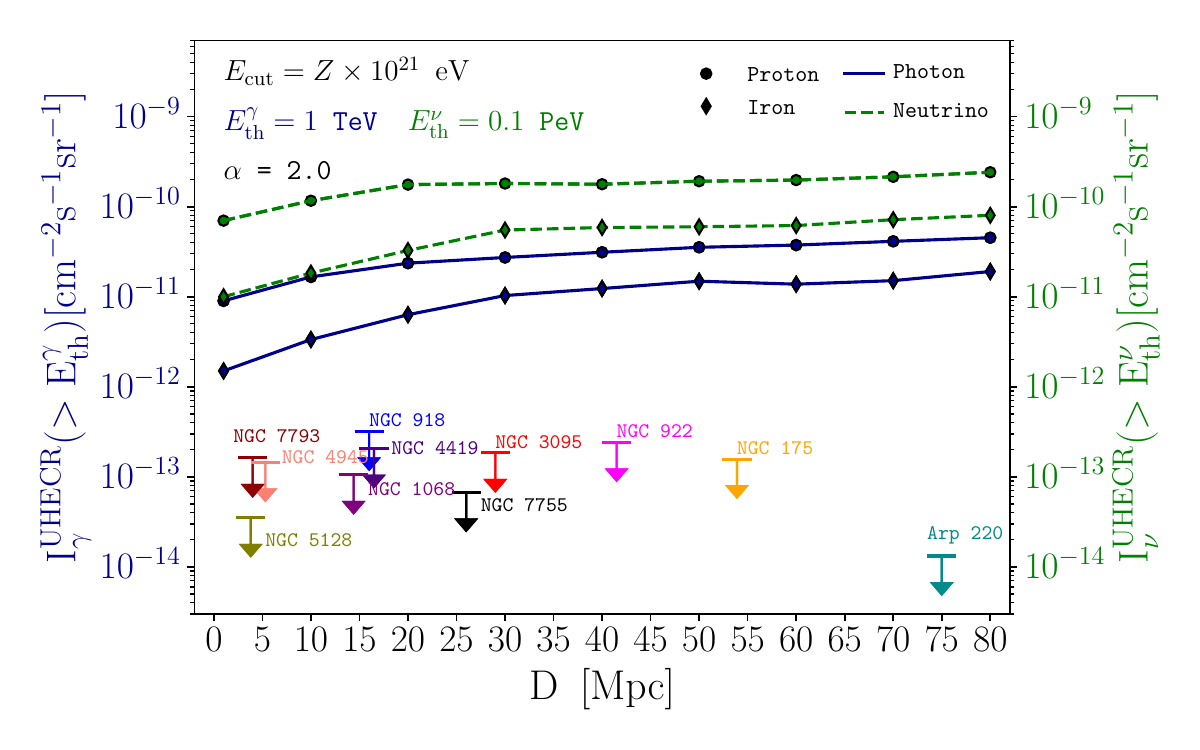}
\caption{
ULs on the integral flux of gamma rays ($I\rm_{\gamma}^{\mathrm{UHECR}}$) and neutrinos ($I\rm_{\nu}^{\mathrm{UHECR}}$) as a function of the source distance, obtained using the UL of the flux from the PAO at 95\% confidence level (CL). The calculations assume a fixed sectral index of $\alpha = 2.0$, a cut-off energy of $E_{\mathrm{cut}} = Z \times 10^{21} \ \mathrm{eV}$, a mixed composition at the source, and energy thresholds of $ E_{\mathrm{th}}^{\gamma} = 1 \ \mathrm{TeV}$ for gamma rays and $ E_{\mathrm{th}}^{\nu} = 0.1 \ \mathrm{PeV}$ for neutrinos. Different marker styles indicate the composition used in each simulation: circles represent protons, while thin diamonds correspond to an iron composition injected at the source. The blue vertical axis represents the integral of the simulated gamma-ray flux as a function of distance (horizontal axis, in megaparsecs), while the green vertical axis represents the integral of the simulated neutrino flux, also as a function of distance. The colored arrows indicate the ULs of the gamma-ray fluxes obtained by the observatories. The solid lines correspond to the simulated photon flux integrals, whereas the dashed lines depict the neutrino flux integrals.}
\label{upperlimits}
\end{figure*}}

\subsection{Upper Limits on the cosmic-ray Luminosity of Individual Sources}\label{individualUL}

We present the effective ULs on cosmic-ray luminosity for eleven previously discussed sources. The maximum UHECR luminosity was determined by integrating the measurements using the method described in Section~\ref{section2}. This approach enables the calculation of an UL on cosmic-ray luminosity for each source, based on GeV-TeV gamma-ray observations.  The method operates under the assumption that UHECRs generate gamma rays along their propagation from the source to Earth. This secondary gamma-ray contribution plays a crucial role in interpreting the total observed flux, providing insights into UHECR interactions and energy losses.  Determining the cosmic-ray luminosity of individual sources is fundamental to constraining cosmic-ray acceleration models and understanding the mechanisms driving particle acceleration in astrophysical environments. These results contribute to the broader effort of characterizing UHECR sources and their role in multimessenger astrophysics.

Figure~\ref{fig:uluhecr} presents the effective UL on cosmic-ray luminosity as a function of the spectral index ($\alpha$) for each selected source. These results were obtained using the composition models defined in Table~\ref{tab:1}, along with the energy thresholds specified by the respective observatories for each source. Additionally, an injection energy cutoff $ E_{\mathrm{cut}} = Z \times 10^{21} \ \mathrm{eV}$ was applied to the injected particles. The results reveal a significant variation in UHECR luminosity depending on the source and spectral characteristics. The highest UL on UHECR luminosity is observed for Arp 220, reaching $1.3 \times 10^{44}$ erg/s, particularly for higher spectral indices (softer spectra). In contrast, the lowest UL is found for NGC 5128, with values around $10^{38}$ erg/s, indicating a substantially lower estimated UHECR luminosity.  

The ULs on cosmic-ray luminosity generally increase as the spectral index $\alpha$ increases, reflecting the reduced contribution of higher-energy cosmic rays in softer spectra. This trend is consistent across all sources, with harder spectra ($\alpha \approx 2.0$) leading to lower luminosity ULs, while softer spectra ($\alpha \approx 3.0$) yield higher luminosity constraints. The variation in ULs also depends on the source composition as indicated by different markers representing mixed cosmic-ray compositions.  Among the sources analyzed, NGC 7755, NGC 922, and NGC 918 exhibit higher UHECR luminosities suggesting that these sources may have a more efficient cosmic-ray acceleration mechanism. In contrast, sources such as NGC 4945 and NGC 5128 present the lowest ULs, implying weaker acceleration conditions or a more substantial attenuation of cosmic rays.

Our method yields an UL on UHECR luminosity of $10^{39}$ erg/s for NGC 1068 at an energy threshold of 200 GeV, based on the gamma-ray flux constraints set by MAGIC \citep{2019ApJ...883..135A}. This value is significantly lower than the $10^{43}$ erg/s estimate reported by \cite{2024ApJ...972...44D}, which is based on gamma-ray observations in the 10–30 TeV range. Several factors contribute to this difference: Their estimate probes higher-energy cosmic rays, which may indicate a more efficient acceleration mechanism in NGC 1068; gamma-ray absorption effects at higher TeV energies require a larger UHECR power to explain the observed flux, as gamma rays interact with photon backgrounds such as the CMB and EBL; their model may assume a different cosmic-ray composition or a scenario where hadronic interactions dominate neutrino production, leading to a higher required UHECR luminosity; finally, differences in how EGMF and cosmic-ray propagation are treated could affect the inferred UHECR power, as UHECR deflections and energy losses impact gamma-ray and neutrino flux predictions. Our results provide complementary constraints in a different energy regime, reinforcing the importance of multimessenger approaches.

The ULs derived in this study provide valuable constraints on the potential of individual sources to accelerate UHECRs and contribute to the observed multimessenger signals. These results emphasize the critical influence of spectral assumptions and nuclear composition models on the inferred CR luminosities, underscoring the importance of continued observational efforts in gamma-ray and neutrino astronomy.
Moreover, the analysis demonstrates the significant impact of magnetic field effects on UHECR propagation. The inclusion of both extragalactic and galactic magnetic fields leads to a reduced fraction of cosmic rays that successfully reach the observer, thereby increasing the inferred upper limits on source luminosities. This conservative introduced through the efficiency factors $\xi_{\mathrm{EGMF}}$ and $\xi_{\mathrm{GMF}}$ ensures that only particles capable of arriving at Earth are considered in the luminosity estimates. As a result, the derived values of $L_{\mathrm{CR}}^{\mathrm{eff}}$ offer a more realistic and physically motivated assessment of the CR luminosity from each source, accounting for magnetic deflections along the entire propagation path.

\begin{figure*}[htbp!]
    \centering
    \begin{minipage}{0.48\textwidth}
        \centering
        \includegraphics[width=\textwidth]{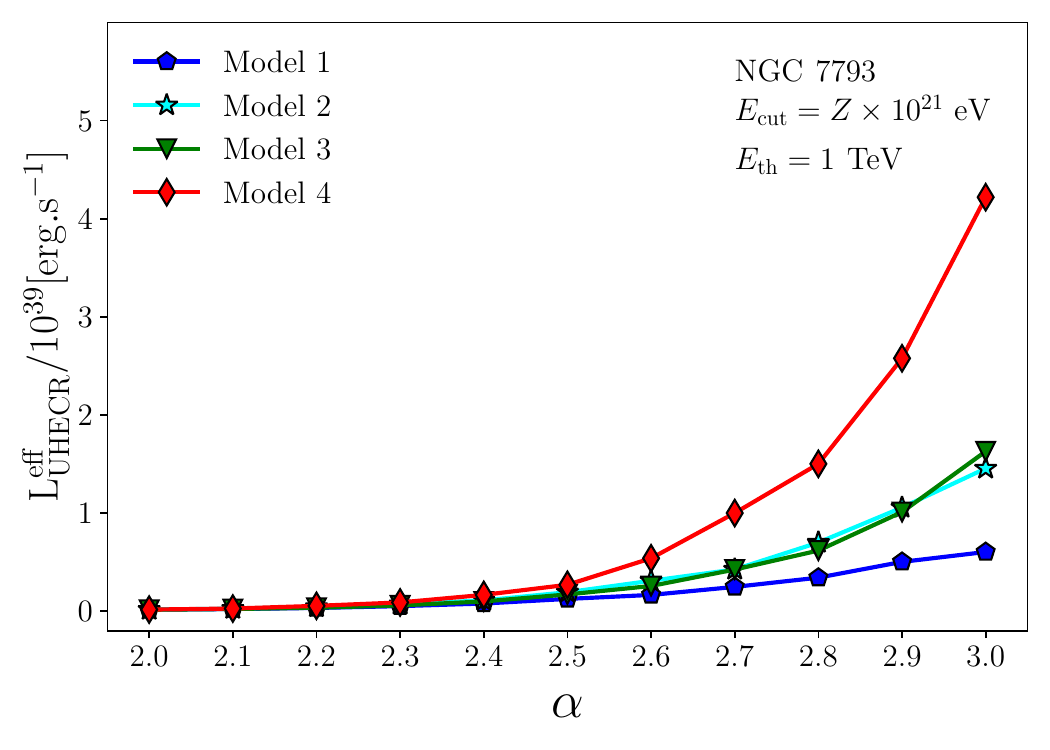}
        \par\vspace{5pt}
        (a) NGC 7793
    \end{minipage}
    \hfill
    \begin{minipage}{0.48\textwidth}
        \centering
        \includegraphics[width=\textwidth]{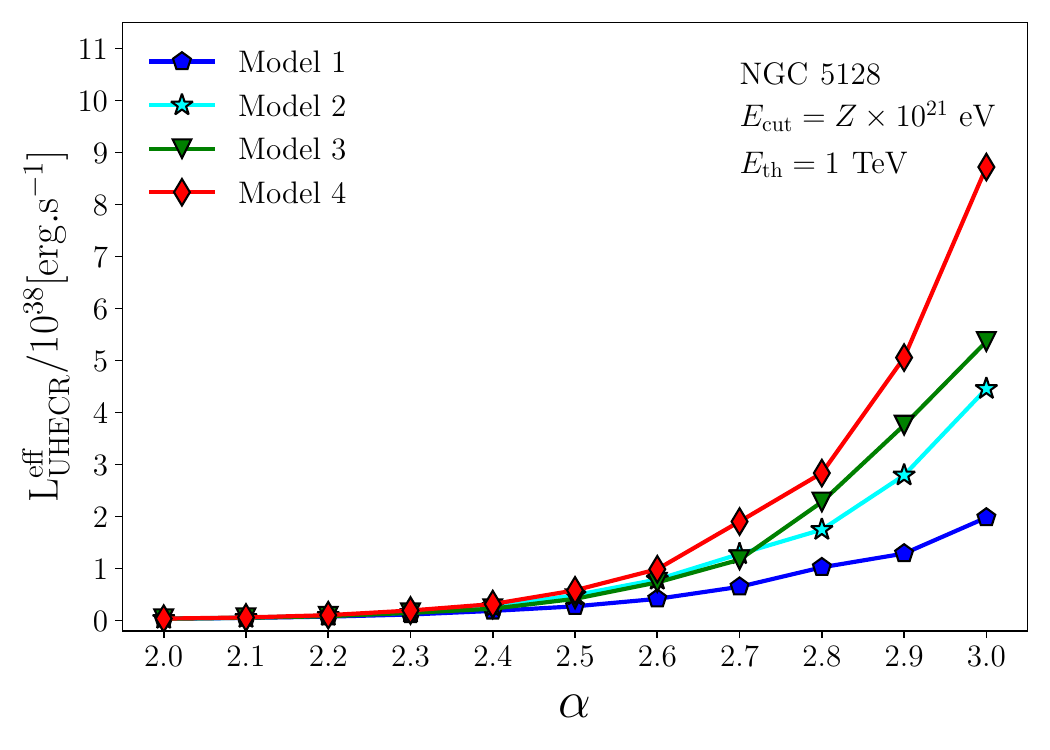}
        \par\vspace{5pt}
        (b) NGC 5128
    \end{minipage}

    \begin{minipage}{0.48\textwidth}
        \centering
        \includegraphics[width=\textwidth]{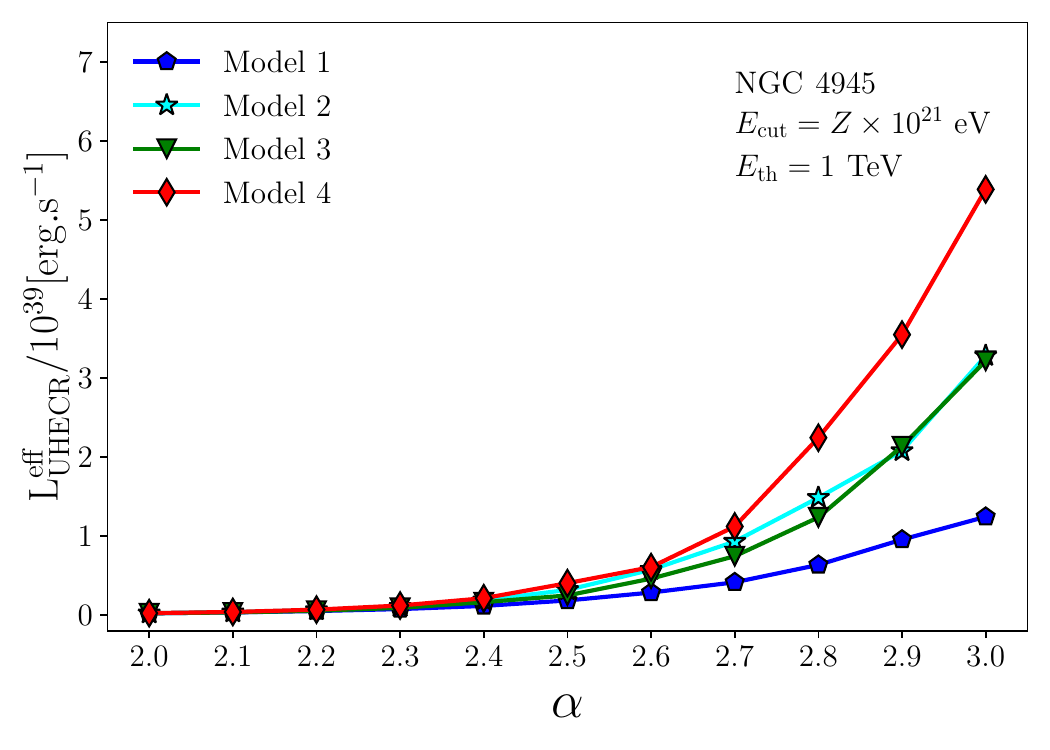}
        \par\vspace{5pt}
        (c) NGC 4945
    \end{minipage}
    \hfill
    \begin{minipage}{0.48\textwidth}
        \centering
        \includegraphics[width=\textwidth]{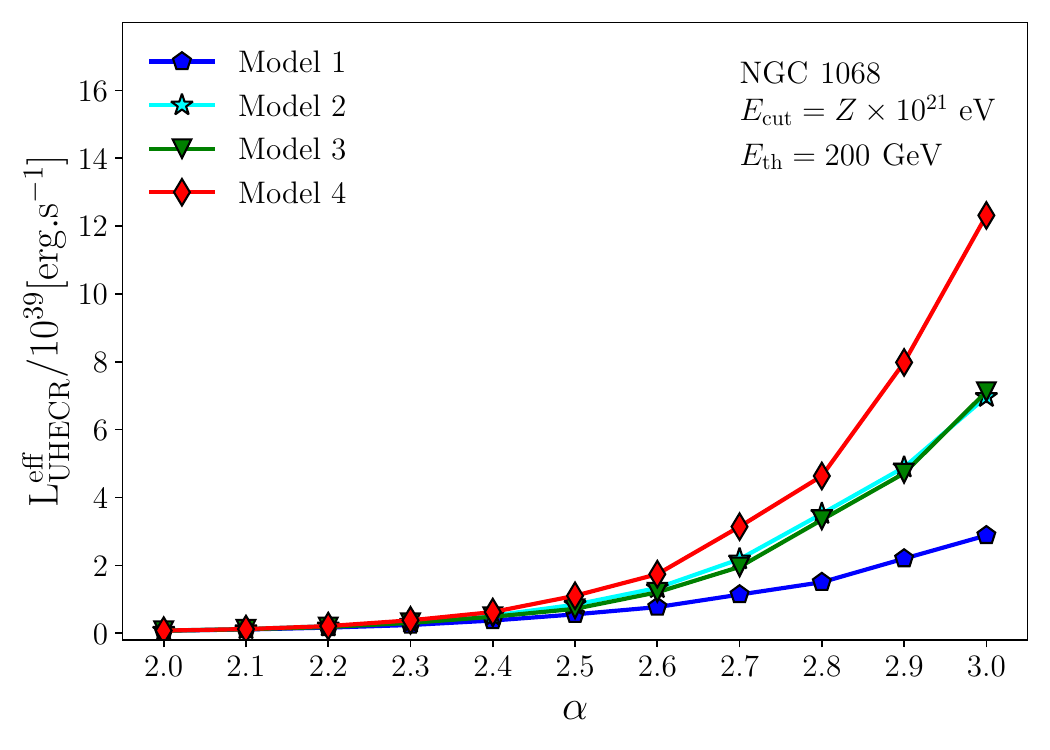}
        \par\vspace{5pt}
        (d) NGC 1068
    \end{minipage}

    \begin{minipage}{0.48\textwidth}
        \centering
        \includegraphics[width=\textwidth]{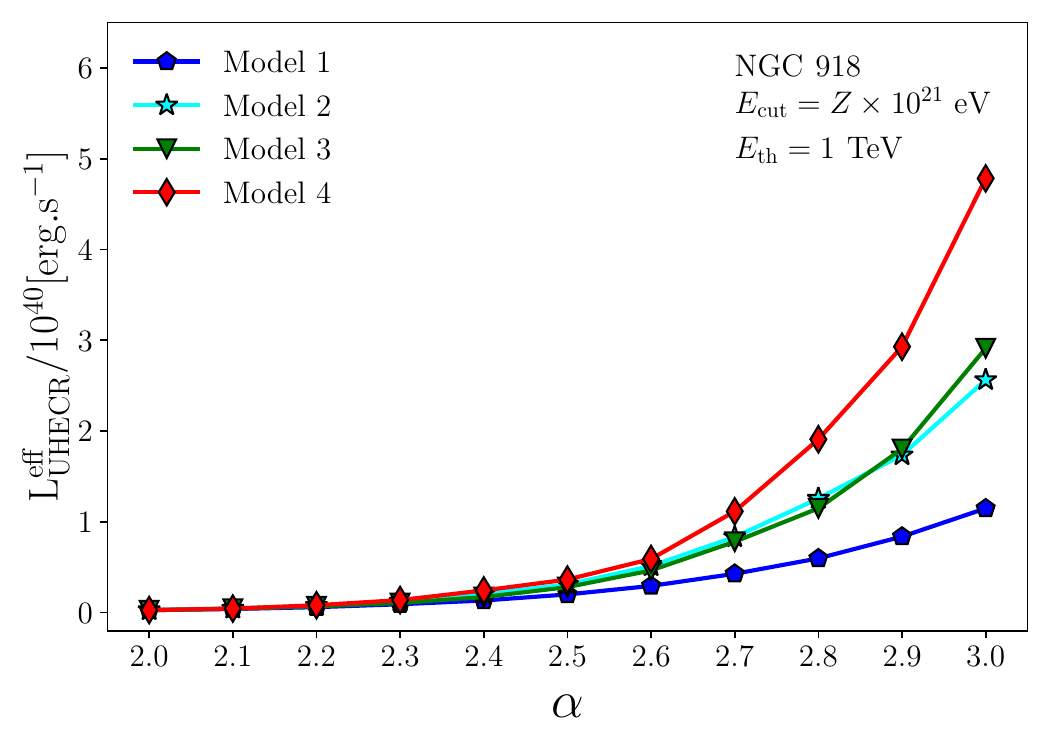}
        \par\vspace{5pt}
        (e) NGC 918
    \end{minipage}
    \hfill
    \begin{minipage}{0.48\textwidth}
        \centering
        \includegraphics[width=\textwidth]{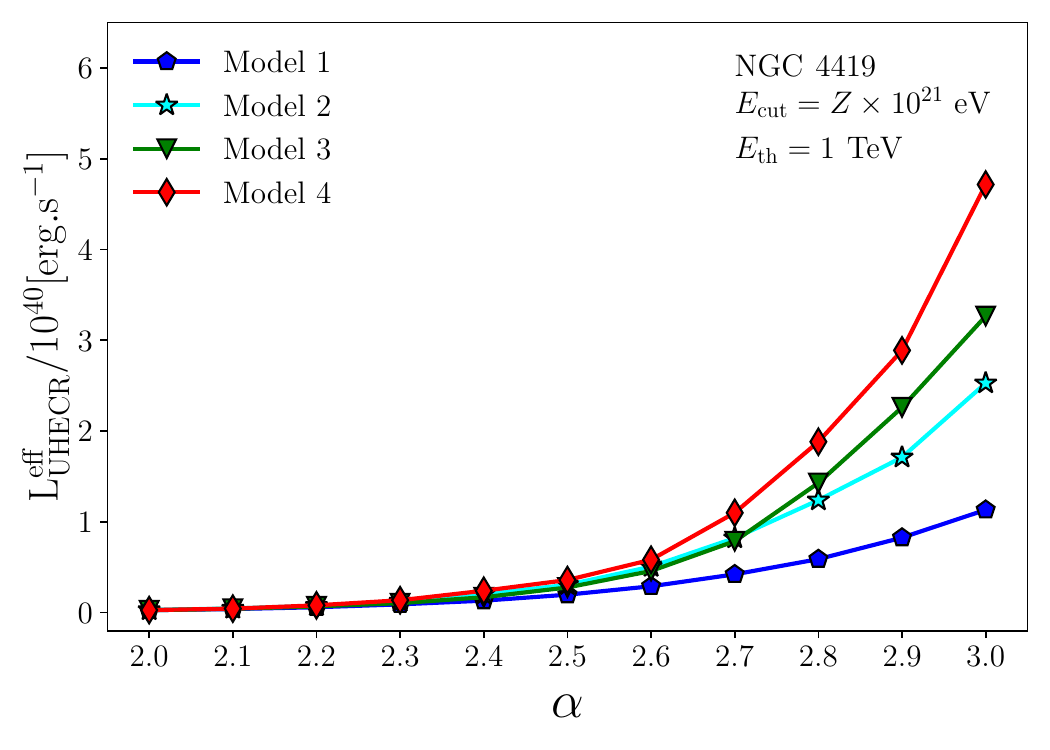}
        \par\vspace{5pt}
        (f) NGC 4419
    \end{minipage}
  
\end{figure*}
\begin{figure*}
\centering
    \begin{minipage}{0.48\textwidth}
        \centering
        \includegraphics[width=\textwidth]{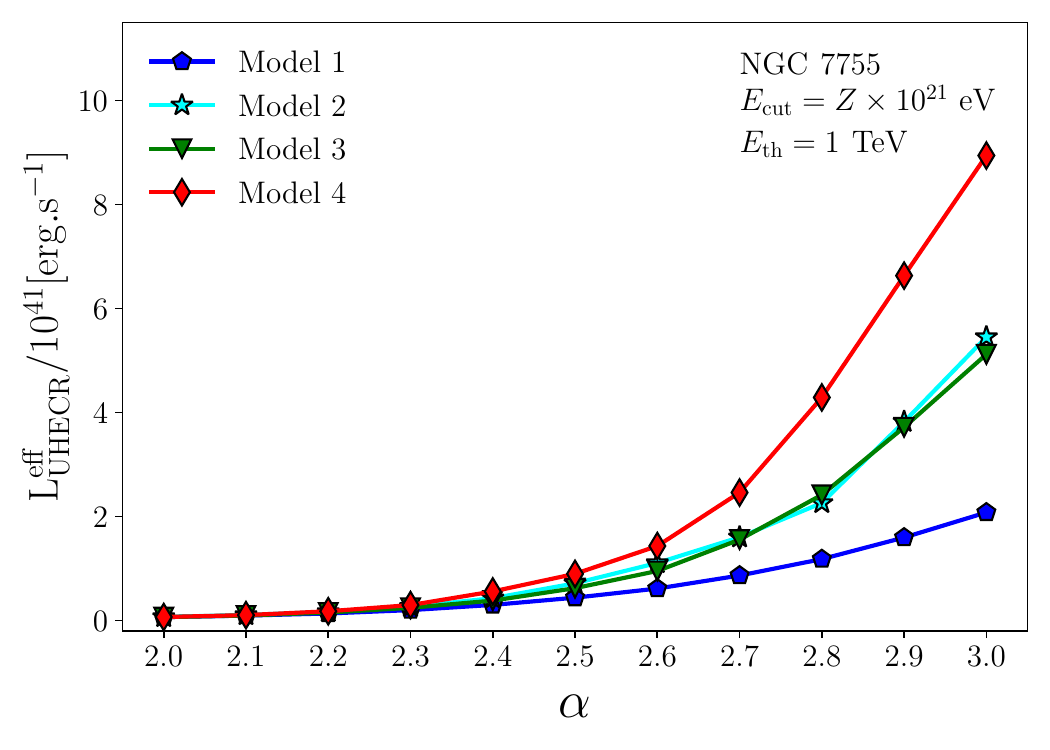}
        \par\vspace{5pt}
        (g) NGC 7755
    \end{minipage}
    \hfill
    \begin{minipage}{0.48\textwidth}
        \centering
        \includegraphics[width=\textwidth]{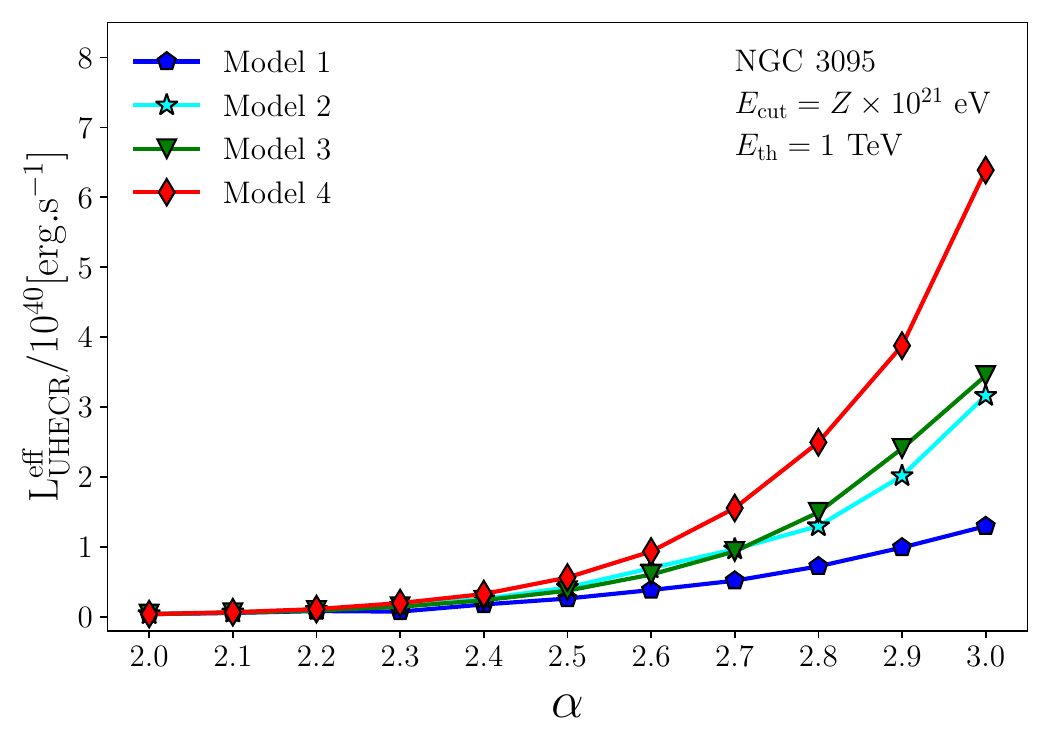}
        \par\vspace{5pt}
        (h) NGC 3095
    \end{minipage}
    \begin{minipage}{0.48\textwidth}
        \centering
        \includegraphics[width=\textwidth]{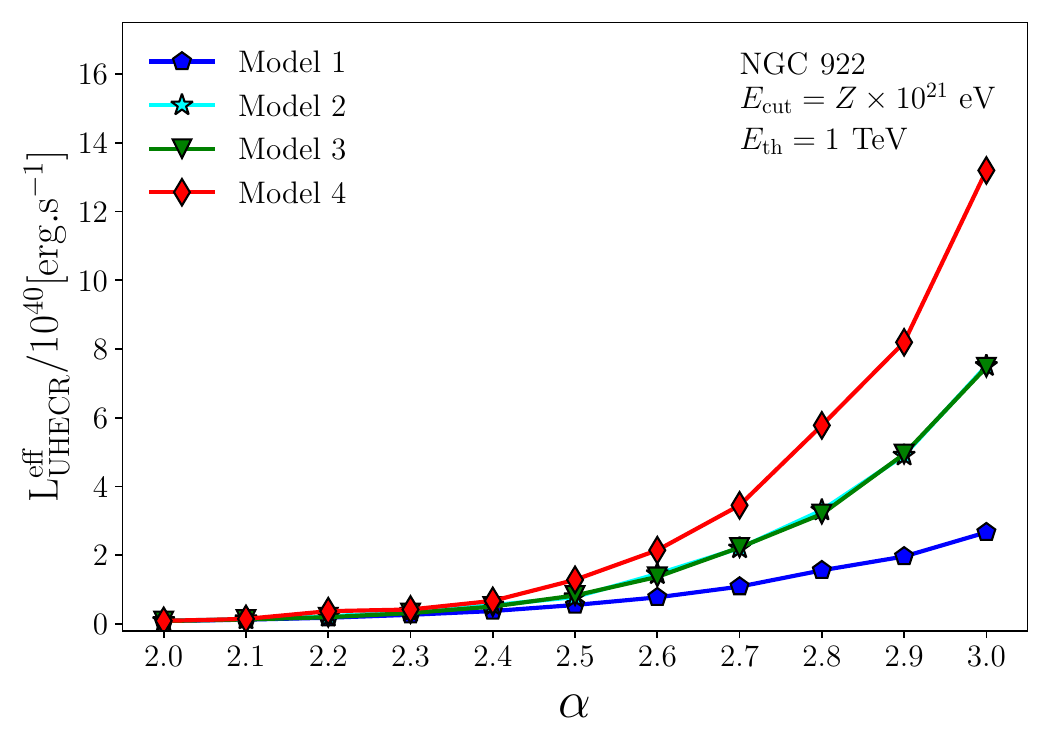}
        \par\vspace{5pt}
        (i) NGC 922
    \end{minipage}
    \hfill
    \begin{minipage}{0.48\textwidth}
        \centering
        \includegraphics[width=\textwidth]{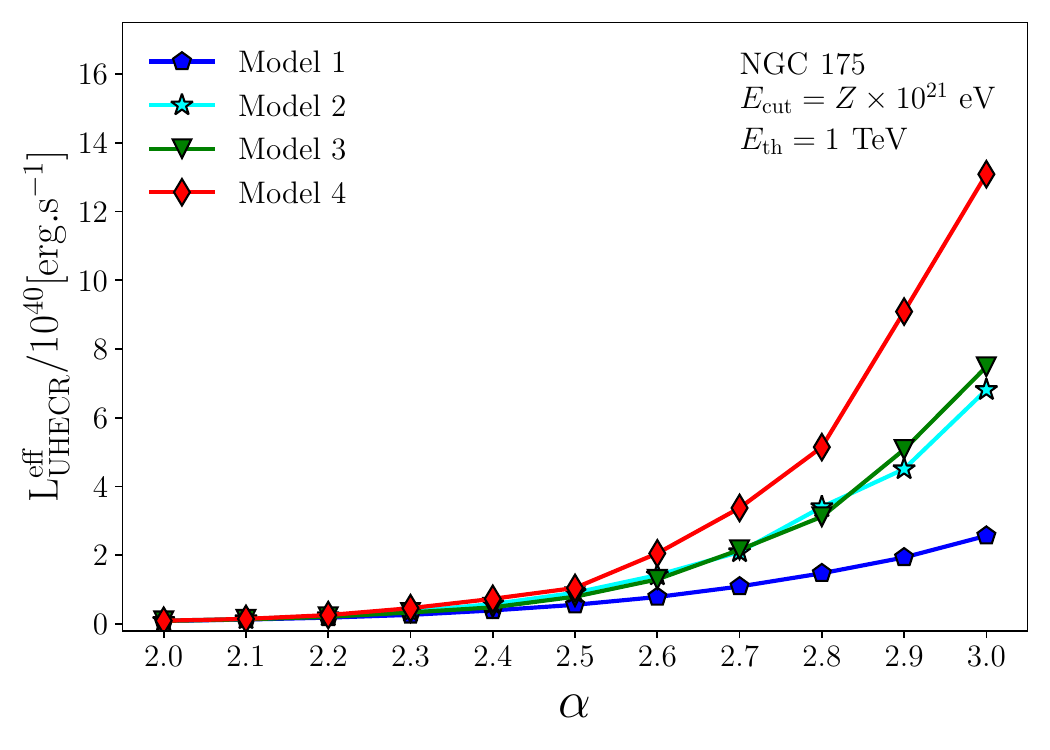}
        \par\vspace{5pt}
        (j) NGC 175
    \end{minipage}

    \begin{minipage}{0.48\textwidth}
        \centering
        \includegraphics[width=\textwidth]{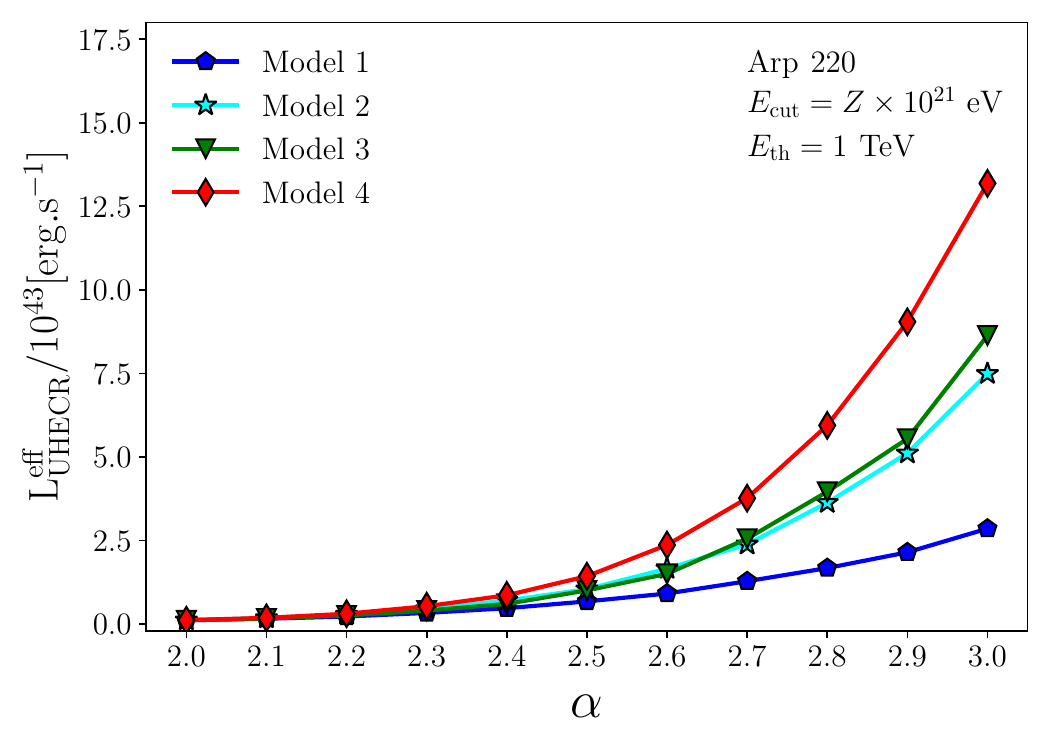}
        \par\vspace{5pt}
        (k) Arp 220
    \end{minipage}

    \caption{ULs on the cosmic-ray luminosity for mixed compositions, as inferred from gamma-ray observations, for different sources as a function of the spectral index ($\alpha$). The ULs are calculated for the specific energy threshold ($E_{\mathrm{th}}$) defined for each source and assuming a fixed cutoff energy of $E_{\mathrm{cut}} = Z \times 10^{21} \ \mathrm{eV}$. The numbered models correspond to the mixed compositions described in Table~\ref{tab:1}.}
    \label{fig:uluhecr}
\end{figure*}

\clearpage
\section{Performance of the CTAO in Observing NGC 1068}\label{sec:ctao}

In addition to the analysis presented in Sections~\ref{section2} and~\ref{section3}, this section investigates whether the enhanced sensitivity of CTAO can lead to stronger constraints on the UHECR luminosity of NGC 1068. While Section~\ref{section3} derives current upper limits based on existing gamma-ray observations, a future detection or a more stringent upper limit from CTAO would directly refine the constraint expressed in Equation~\ref{eq:LCR}. In this context, we examine the projected gamma-ray spectrum from NGC 1068 and assess whether the anticipated performance of CTAO can reach flux levels capable of improving or validating the previously established luminosity bounds.

This analysis employs methodologies developed in~\cite{Costa2024} and further refined by~\cite{Sousa_2025} and~\cite{Sasse_2025}, to evaluate CTAO’s potential to detect gamma-ray emission from NGC 1068 in the multi-TeV energy range. Furthermore, we explore how these observations can complement our UL constraints on the UHECR luminosity. As the next-generation ground-based gamma-ray observatory, the CTAO is designed to significantly enhance our ability to detect VHE gamma rays \citep{2019scta.book.....C}. Its increased sensitivity and superior angular resolution will provide deeper insights into the multimessenger signals associated with UHECRs. NGC 1068 has been identified as a high-energy neutrino source by IceCube \citep{doi:10.1126/science.abg3395} and is considered a promising candidate for hadronic interactions in AGNs. Studying its gamma-ray emission is therefore crucial for understanding the underlying physical processes.

To characterize the gamma-ray emission centered at the NGC 1068 position ($\mathrm{R.A.} = 40.669^\circ, \mathrm{Dec.} = -0.013^\circ$)\footnote{Coordinates are given in the International Celestial Reference System (ICRS) using Right Ascension (R.A.) and Declination (Dec.).}, we constructed a source spectral model spanning the multi-GeV to multi-TeV energy range. This is achieved through a joint likelihood fit of data from previous observations of electromagnetic counterparts within a radius of $< 0.11^{\circ}$. For each counterpart included in the joint-likelihood analysis, the published datasets and their bibliographic sources are summarized in Table~4. The intrinsic spectral model of the source is defined by a power law:  
\begin{equation}\label{eq:int_model}
    \Phi(E) = \Phi_0 \left(\frac{E}{E_0} \right)^{- \alpha},
\end{equation}
where $\Phi_{0}$ is the amplitude, $ E_0 = 1 \ \mathrm{TeV}$ is the reference energy, and $\alpha$ is the spectral index. To account for absorption, the absorbed spectrum is given by:
\begin{equation}\label{eq:abs_model}
    \Phi (E)_{\mathrm{abs}} = \Phi (E) \exp^{-\tau (E, z)},
\end{equation}
where $\tau$ is the optical depth, and $z$ is the source redshift. The absorption is based on the Saldana-Lopez EBL model~\citep{Saldana_Lopez_2021}.  
Table~\ref{tab:skymodel} presents the celestial coordinates, redshift, and spectral model parameters, obtained by likelihood fit, for the modeled source. It includes the spectral model parameters, derived from the simulated CTAO observation, described in detail below. Moreover, Table~\ref{tab:counterparts} provides the celestial coordinates and angular separations from the region's center for the counterparts included in the likelihood fit.
\begin{table}[htbp!]
    \begin{minipage}{0.47\textwidth}
    \centering
     \caption{Celestial coordinates, given in the International Celestial Reference System (ICRS) using Right Ascension (R.A.) and Declination (Dec.), and redshift $z$ for the NGC 1068 region. The gamma-ray spectra were modeled using a power law with extragalactic background light (EBL) absorption, $\Phi(E) = \Phi_{0} (E/E_{0})^{- \alpha}\exp^{-\tau (E, z)}$, where $\tau$ is the optical depth from the Saldana-Lopez model \cite{Saldana_Lopez_2021}. The spectral model parameters  include amplitude $\Phi_{0}$ (scaled by $10^{-14}$) and spectral index $\alpha$.}\label{tab:skymodel}
        \begin{tabular}{ccc}
        \toprule 
        \centering
          \textbf{Position} & \multicolumn{2}{c}{$(\mathrm{R.A.}=40.669^\circ, \mathrm{Dec.}=-0.013^\circ)^{a}$}  \\ 
         \textbf{Redshift $z$} & \multicolumn{2}{c}{$0.003793$$^{b}$}  \\ 
              \textbf{absorption} & \multicolumn{2}{c}{Saldana-Lopez EBL model$^{c}$}  \\
         \textbf{fit} & \textbf{source model} & \textbf{CTAO South (200h)} \\
            \cmidrule(l){2-3} 
         $\Phi_0 \times 10^{-14}$           \\ 
        ($\mathrm{cm^{-2}\,s^{-1}\,TeV^{-1}}$) & $6.334 \pm 2.421$ & $6.043 \pm 0.504$ \\
         $\alpha$ & 
        $2.36 \pm 0.06$ &$2.38 \pm 0.06$ \\
         \bottomrule
        \end{tabular}
    \tablerefs{$^a$\cite{Spinoglio_2022}, $^b$\cite{Huchra1999}, $^c$\cite{Saldana_Lopez_2021}}
     \end{minipage}
\end{table}

\begin{table}[htbp!]
    \begin{minipage}{0.47\textwidth}
    \centering
     \caption{Celestial coordinates, given in the ICRS using R.A. and Dec., as well as the angular separation (Sep.) from the region's center (at the NGC 1068 position ($\mathrm{R.A.} = 40.669^\circ, \mathrm{Dec.} = -0.013^\circ$)~\citep{Spinoglio_2022}) for the counterparts included in the simultaneous likelihood fit.}
    \label{tab:counterparts}
        \begin{tabular}{llll}
        \toprule
        \textbf{Counterparts} & \textbf{R.A.} & \textbf{Dec.} & \textbf{Sep.} \\
        \midrule
        3FHL J0242.7-0002$^{a}$  & $40.676^\circ$ & $-0.043^\circ$ & $0.030^\circ$ \\ 
        3FGL J0242.7-0001$^{b}$  & $40.680^\circ$ & $-0.026^\circ$ & $0.016^\circ$ \\
        4FGL J0242.6-0000$^{c,d}$  & $40.667^\circ$ & $-0.007^\circ$ & $0.007^\circ$ \\
        MAGIC$^{e, f}$  & $40.669^\circ$ & $-0.013^\circ$  & - \\
         \bottomrule
        \end{tabular}
    \tablerefs{$^a$\cite{Ajello_2017}, $^b$\cite{Acero_2015}, $^c$\cite{Ballet_2023}, $^d$\cite{Abdollahi_2022}, $^e$\cite{Spinoglio_2022}, $^f$\cite{2019ApJ...883..135A}}
     \end{minipage}
\end{table}

The expected CTAO's observational performance is analyzed using the 1D ON/OFF method described in~\cite{Piano_2022} with the background (OFF region) determined using the ``Alpha Configuration'' Instrument Response Functions (IRFs) for the southern array provided by the CTA Consortium and CTAO (version prod5 v0.1~\cite{CTAOIRFS})\footnote{In the ``Alpha Configuration'', the IRFs for the southern array include: CTAO South with 14 Medium-Sized Telescopes (MSTs) and 37 Small-Sized Telescopes (SSTs); CTAO South-MSTs with 14 MSTs; and CTAO South-SSTs with 37 SSTs.  The IRFs are optimized for a 50-hour observation time at a zenith angle of $20^\circ$, pointing azimuth-averaged.}. The ON region is a circular area with a radius of $0.11^\circ$, centered on the source position. We impose a signal/background ratio of at least $1/20$. Observations were conducted in wobbling mode, with a $0.5^\circ$ offset from the source position, assuming parallel telescope pointing. In this study, we produced the differential sensitivity curves, utilizing a constant flux enclosure fraction of $68\%$ to address the energy-dependent region size and enhance the background estimation. Additionally, a minimum of five expected signal counts per bin and a significance threshold of $3\sigma$ per bin were imposed. Furthermore, the gamma-ray excess was required to exceed $10\%$ of the background count. The flux points of the CTAO's observation were generated in the energy range of $0.1-32\ \mathrm{TeV}$. All analyzes were carried out using the Gammapy\footnote{\url{https://gammapy.org/}} package \cite{Donath_2015, Donath_2023}.

\begin{figure}[htb]
    \centering
    \begin{minipage}{0.48\textwidth}
        \centering
        \includegraphics[width=\textwidth]{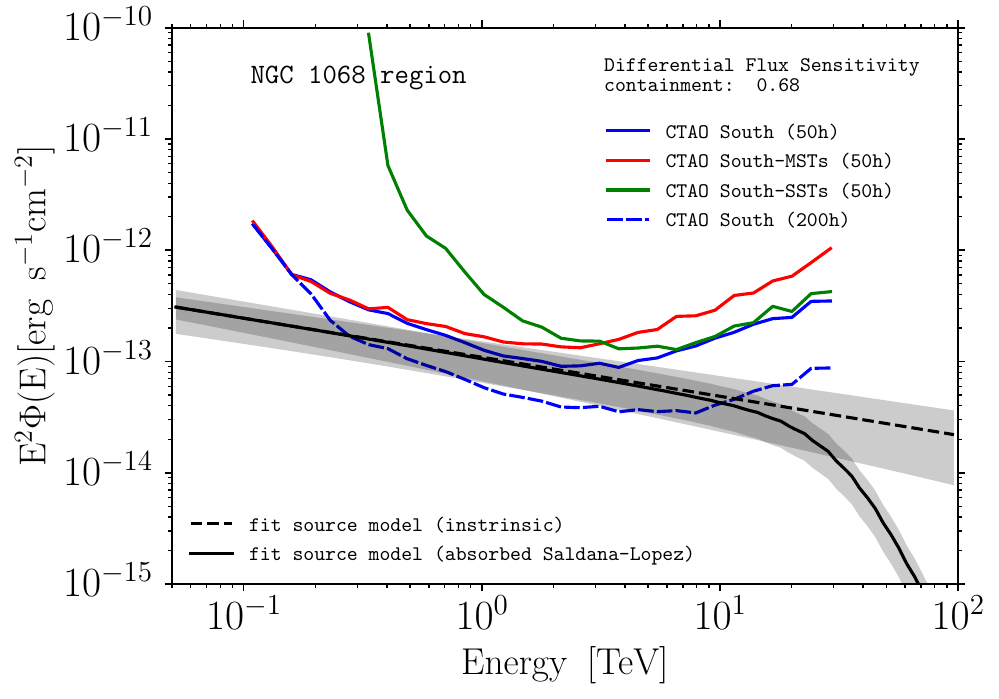}
        \par\vspace{5pt}
        (a) CTAO Sensitivity Curves.
    \end{minipage}
    \begin{minipage}{0.48\textwidth} 
        \centering
        \includegraphics[width=\textwidth]{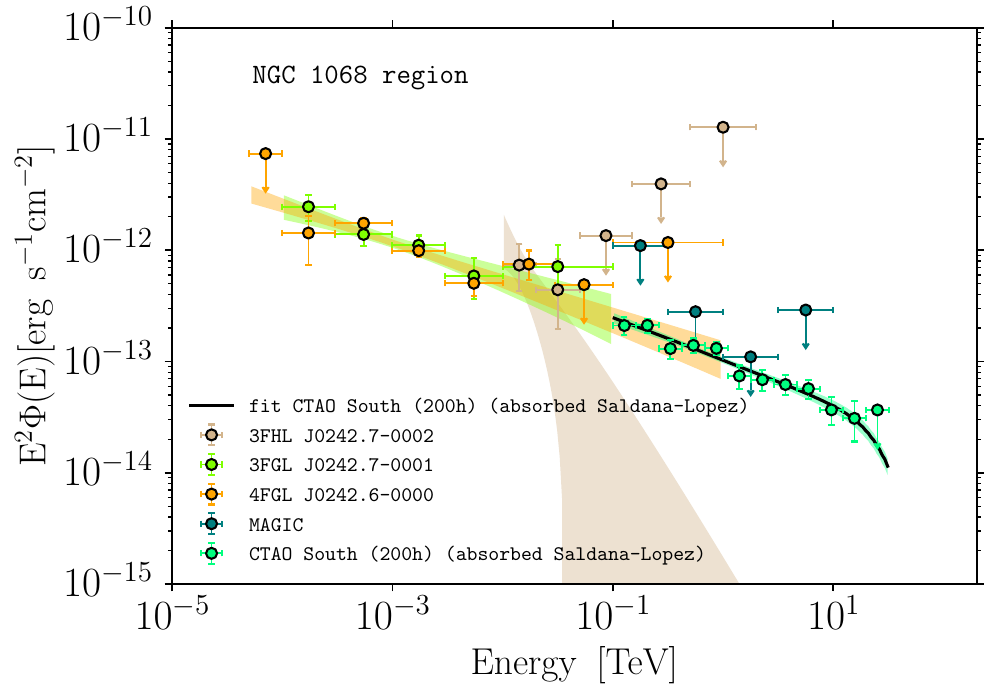} 
        \par\vspace{5pt} 
        (b) Modeled gamma-ray spectra for NGC 1068
    \end{minipage}
    \hfill 
\caption{Energy-dependent differential flux sensitivity and spectral energy distribution (SED) for the NGC 1068 region. (a) The black dashed line represents the intrinsic spectral model (Equation~\ref{eq:int_model}), while the black solid line includes attenuation due to extragalactic background light (EBL), modeled using Saldana-Lopez~\citep{Saldana_Lopez_2021} (Equation~\ref{eq:abs_model}). Model parameters are in Table~\ref{tab:skymodel}, with shaded regions indicating errors. Sensitivity curves for CTAO South are shown with different configurations: solid blue (14 MSTs + 37 SSTs), solid red (MSTs only), and solid green (SSTs only) for 50-hour observations; the dashed blue line represents the full array for 200 hours. IRFs are optimized for a 50-hour exposure at a $20^\circ$ zenith angle with azimuth-averaged pointing. (b) Modeled gamma-ray flux for NGC 1068 overlaid with Fermi-LAT (3FHL, 3FGL, 4FGL) and MAGIC data (Table~\ref{tab:counterparts}). The black solid line represents the absorbed spectral model from Saldana-Lopez~\citep{Saldana_Lopez_2021} (Table~\ref{tab:skymodel}), while lighter green circles show CTAO South flux points for a 50-hour observation. Error bars indicate statistical uncertainties.}
    \label{fig:ctao_sensitivity}
\end{figure}

Figure~\ref{fig:ctao_sensitivity}-a presents the energy-dependent differential flux sensitivity and spectral model predictions for the NGC 1068 region, evaluated in the context of CTAO observations. The solid-colored curves represent the sensitivity limits for different CTAO southern array configurations and observation times. These configurations demonstrate CTAO’s ability to detect gamma-ray fluxes in the multi-TeV range with optimized IRFs at a $20^{\circ}$ zenith angle for a 200-hour observation time. The black solid line shows the intrinsic spectral model of NGC 1068, while the black dashed line represents the absorbed spectral model, which accounts for the EBL attenuation based on the Saldana-Lopez model~\cite{Saldana_Lopez_2021}. At higher energies, absorption significantly attenuates the gamma-ray flux due to photon-photon interactions with extragalactic background light (EBL). The comparison between these two spectral models highlights the impact of EBL absorption on the detectability of gamma rays from AGN environments. Additionally, Figure~\ref{fig:ctao_sensitivity}-b includes observed flux data points from MAGIC and Fermi-LAT (3FGL and 4FGL). The plot illustrates the importance of CTAO’s extended sensitivity in the TeV range, where previous instruments have set only upper limits. The ability of CTAO to probe deeper into this energy range will enhance our ability to constrain the physical mechanisms driving AGN activity and their role as potential UHECR accelerators.

In summary, Figure~\ref{fig:ctao_sensitivity} underscores the critical role of CTAO in the multimessenger study of NGC 1068. By significantly improving the sensitivity to VHE gamma rays, CTAO can provide essential constraints on UHECR-induced emissions, complementing existing gamma-ray and neutrino observations. This analysis strengthens the connection between gamma-ray, neutrino, and cosmic-ray physics, advancing our understanding of AGNs as cosmic-ray accelerators.

\section{Conclusions}\label{conclusions}

This study presents a detailed analysis of secondary fluxes generated by the propagation of UHECRs, focusing on the production of gamma rays and neutrinos through non-thermal processes. Using extragalactic simulations with CRPropa3~\citep{Alves_Batista_2022}, we modeled the expected multimessenger fluxes and evaluated their detectability. The key results of our analyses are summarized below.

The first part of this study examined the integral fluxes of gamma rays, evaluating the impact of various parameters, including cutoff energy $ E_{\mathrm{cut}}$, spectral index $\alpha$, and energy threshold $ E_{\mathrm{th}}$. We have updated and extended the results previously presented by \citealt{Supanitsky_2013, Anjos_2014}, incorporating new data and refined modeling approaches. Since the integral flux of neutrinos can be derived using a similar approach, the same parameter study was applied to the neutrino flux predictions. These results are discussed in Section~\ref{section2}. In Section~\ref{2.3}, we establish ULs on neutrino flux for individual sources, with a particular emphasis on NGC 1068, one of the most significant extragalactic neutrino sources identified by IceCube \citep{doi:10.1126/science.aat2890}. Once the UL on UHECR luminosity was derived, it became possible to set corresponding ULs on the neutrino flux, imposing constraints on neutrino production models. 

A detailed analysis of EGMF was performed in Sections~\ref{2.5} and \ref{2.6}, where we explored the impact of different strengths of EGMF on the propagation of UHECR. To ensure that at least 90\% of the injected particles reach the observer, we limit the maximum EGMF intensity to $1 \times 10^{-14} \ \mathrm{G}$. This result plays a crucial role in understanding UHECR deflections and energy losses in extragalactic environments.

In Section~\ref{section3}, we identified eleven sources to determine ULs on their UHECR luminosity. This was achieved using the measured ULs on the integral GeV-TeV gamma-ray flux and modeling the propagation of UHECR particles from their sources to Earth. The results impose stringent constraints on the assumed cosmic-ray acceleration models and reveal fundamental differences in cosmic-ray emission properties across different source types. The ULs for each source were derived from observations by the H.E.S.S. (10 sources) \citep{refId0, RC} and MAGIC (1 source) telescopes \citep{2019ApJ...883..135A}. Our methodology provided an UL of $10^{39} \ \mathrm{erg \ s^{-1}}$ for NGC 1068 at $ E_{\mathrm{th}} = 200$ GeV, based on gamma-ray flux constraints set by MAGIC~\citep{2019ApJ...883..135A}. 

An important outcome of this study is the incorporation of both extragalactic and galactic magnetic field effects in the derivation of upper limits on UHECR luminosity of point-like sources. By introducing the deflection efficiency factors $\xi_{\mathrm{EGMF}}$ and $\xi_{\mathrm{GMF}}$, we account for the fraction of cosmic rays that successfully reach Earth after propagation. This correction leads to a more conservative and physically realistic estimate of source luminosity, as it filters out particles deflected away from the line of sight. These corrections are essential for connecting gamma-ray observations with plausible source energetics.

Finally, in Section~\ref{sec:ctao}, we assessed the capabilities of the CTAO in detecting gamma-ray emissions from NGC 1068. CTAO’s enhanced capability to detect VHE gamma rays enables it to place more stringent limits on UHECR-related emissions, providing complementary insights alongside existing gamma-ray and neutrino measurements. 

This study demonstrates the effectiveness of combining gamma-ray and neutrino constraints to probe UHECR sources and highlights the role of magnetic fields in shaping cosmic-ray propagation. The upper limits on UHECR luminosity impose critical constraints on high-energy astrophysical models, emphasizing the necessity of continued multimessenger observations with next-generation facilities like CTAO \citep{2019scta.book.....C}, KM3NeT \citep{km3net_2025} and others.

\section{Acknowledgments}
We thank the anonymous reviewer for insightful comments and suggestions that substantially improved the clarity of this paper.
This study was financed in part by the Coordenação de Aperfeiçoamento de Pessoal de Nível Superior – Brasil (CAPES) – Finance Code 001. R.S, R. J. C, R.C.A. and C.H.C.-A. acknowledge the financial support from the NAPI “Fenômenos Extremos do Universo” of Fundação de Apoio à Ciência, Tecnologia e Inovação do Paraná. R.C.A. research is supported by CAPES/Alexander von Humboldt Program (88881.800216/2022-01), CNPq (310448/2021-2) and (4000045/2023-0), Araucária Foundation (698/2022) and (721/2022) and FAPESP (2021/01089-1). R.C.A. gratefully acknowledges the Max Planck Institute for Nuclear Physics for their warm hospitality and support during her visit, which provided a conducive environment for fruitful discussions and collaborations. R.C.A. also acknowledges the support of L’Oreal Brazil, with the partnership of ABC and UNESCO in Brazil. The authors express their sincere gratitude to the H.E.S.S. and MAGIC observatories for providing upper limit measurements on gamma-ray fluxes, which played a crucial role in our analysis. Additionally, the authors acknowledge the AWS Cloud Credit/CNPq and the National Laboratory for Scientific Computing (LNCC/MCTI, Brazil) for providing HPC resources through the SDumont supercomputer, which significantly supported the computational aspects of this research, which have contributed to the research results reported in this paper. URL: https://sdumont.lncc.br. The research also used Gammapy, a Python package developed by the community for TeV gamma-ray astronomy \citep{Deil_2017, Donath_2023}, accessible at \href{https://www.gammapy.org}{https://www.gammapy.org}. In addition, we used the instrument response functions for the Cherenkov Telescope Array Observatory (CTAO) provided by the CTA Consortium and CTAO. For detailed information on these instrument response functions, see \href{https://www.ctao-observatory.org/science/cta-performance}{https://www.ctao-observatory.org/science/cta-performance} (version prod5 v0.1; \cite{CTAOIRFS})

\bibliographystyle{aasjournal}



\end{document}